\documentclass[12pt]{article}
\usepackage[T1]{fontenc}
\usepackage[utf8]{inputenc}
\usepackage{authblk}

\title{\LARGE \textbf{COMBO} and \textbf{COMMA}: \texttt{R} packages for regression modeling and inference in the presence of misclassified binary mediator or outcome variables}

\author[1,2]{Kimberly A. Hochstedler Webb}
\author[2]{Martin T. Wells}
\affil[1]{Division of General Internal Medicine, Department of Medicine, University of Pittsburgh, Pittsburgh, PA}
\affil[2]{Department of Statistics and Data Science, Cornell University, Ithaca, NY}

\usepackage[super,sort&compress]{natbib}
\setcitestyle{comma,numbers,super,open={},close={}} 
\usepackage{graphicx}
\usepackage[paperwidth=8.5in,paperheight=11.0in,top=1in, bottom=1in, left=1in, right=1in,lines=25]{geometry}
\usepackage[title,toc,titletoc,page]{appendix}

\usepackage{amsmath}
\usepackage{bbm}
\usepackage{threeparttable}
\usepackage{makecell}

\usepackage[svgnames]{xcolor}
\usepackage{listings}
\usepackage{hyperref}

\usepackage{setspace}
\DeclareMathOperator{\logit}{logit}

\begin{document}

\maketitle

\begin{abstract}
    Misclassified binary outcome or mediator variables can cause unpredictable bias in resulting parameter estimates. As more datasets that were not originally collected for research purposes are being used for studies in the social and health sciences, the need for methods that address data quality concerns is growing. In this paper, we describe two R packages, \textbf{COMBO} and \textbf{COMMA}, that implement bias-correction methods for misclassified binary outcome and mediator variables, respectively. These likelihood-based approaches do not require gold standard measures and allow for estimation of sensitivity and specificity rates for the misclassified variable(s). In addition, these R packages automatically apply crucial label switching corrections, allowing researchers to circumvent the inherent permutation invariance of the misclassification model likelihood. We demonstrate \textbf{COMBO} for single-outcome cases using a study of bar exam passage. We develop and evaluate a risk prediction model based on noisy indicators in a pretrial risk assessment study to demonstrate \textbf{COMBO} for multi-outcome cases. In addition, we use \textbf{COMMA} to evaluate the mediating effect of potentially misdiagnosed gestational hypertension on the maternal ethnicity-birthweight relationship.

    
\end{abstract}







\section[Introduction]{Introduction}

In randomized and observational studies, imperfectly measured data has the potential to bias association estimates in unpredictable ways. Problems arising due to measurement error cannot be easily overcome by increasing sample sizes and collecting more data. Moreover, error-prone datasets that were originally collected for non-research purposes, like electronic health records and administrative data, are increasingly being used for scientific purposes \citep{omalley2005measuring}. Thus, user-friendly statistical methods and software that allow analysts to address potential biases in error-prone datasets are valuable to methodological and applied researchers alike.  

In this paper, we focus specifically on the measurement error problem of ``misclassification" in binary mediator and outcome variables. Binary variable misclassification occurs when an observed two-category variable is incorrectly recorded or measured in a dataset. Depending on the relationship of this misclassified variable with others in the dataset and the nature of the misclassification, varying degrees of bias in parameter estimation may occur. If the misclassification is in an exposure variable or an outcome variable and if that misclassification is random, for example, regression parameter estimates are generally biased toward the null \citep{ziegler2020binary}. If exposure and outcome variable misclassification depends on other variables in the dataset, on the other hand, resulting regression parameters may flip sides or be exaggerated \citep{nguimkeu2021regression, grace2016statistical}. If \textit{covariate-related} outcome variable misclassification is suspected, the resulting bias can be difficult to overcome \citep{beesley2020statistical, zhang2020genetic}.

In cases of covariate-dependent outcome misclassification, the likelihood of binary outcome misclassification is associated with predictor variables in the dataset. Neuhaus (1999)\cite{neuhaus1999bias} developed analytic expressions of the efficiency of estimators that utilize error-prone binary outcome variables, illustrating the loss of information and substantial bias that misclassified outcome variables can cause in an analysis. Existing methods to correct this bias often require a complete or partial validation study, in which the error-prone variable is audited using a gold standard instrument or data source and the resulting misclassification rates are used to correct the analysis \citep{lotspeich2021using, lotspeich2022efficient, tang2015binary}. Naturally, these methods fail when such a gold standard is not available or simply does not exist. In such cases, analysts might conduct a study with known misclassification rates or a sensitivity analysis with a range of plausible misclassification rates \citep{lyles2010sensitivity, nguimkeu2021regression, magder1997logistic, valeri2014estimation}. These methods are attractive due to their ease of implementation. However, these approaches are difficult to scale when a complex relationship is suspected between the misclassified variable and other predictors in the dataset, as in cases of covariate-dependent misclassification.  

Webb and Wells (2024)\cite{webb2024effect}, Webb and Wells (2023)\cite{webb2023statistical}, and Webb, Riley, and Wells (2023)\cite{webb2023assessment} propose likelihood-based approaches to overcome bias caused by misclassified mediator or outcome variables in association studies. These approaches incorporate covariate-dependence of a misclassified variable and enable the estimation of misclassification rates. Such methods are useful in cases where a gold standard is not available, making validation studies impossible, and where a sensitivity analysis would be infeasible. The methods introduced in Webb and Wells (2024)\cite{webb2024effect}, Webb and Wells (2023)\cite{webb2023statistical}, and Webb, Riley, and Wells (2023)\cite{webb2023assessment} build on that of Beesley and Mukherjee (2020)\cite{beesley2020statistical}, who introduce an approach to correct for covariate-dependent misclassification in outcome variable that is measured with imperfect sensitivity and perfect specificity. The work in Webb and Wells (2023)\cite{webb2023statistical} relaxes the perfect specificity assumption, allowing both false positive and false negative misclassification, and this approach is adopted for extensions in Webb and Wells (2024)\cite{webb2024effect} and Webb, Riley, and Wells (2023)\cite{webb2023assessment}.

In this paper, we describe two R packages, \textbf{COMBO} and \textbf{COMMA}. \textbf{COMBO} implements methods proposed in Webb and Wells (2023)\cite{webb2023statistical} and  Webb, Riley, and Wells (2023)\cite{webb2023assessment} to correct for bias in association studies caused by misclassified binary outcome variables and multiple, sequential misclassified binary outcome variables, respectively. \textbf{COMMA} implements three methods proposed in Webb and Wells (2024)\cite{webb2024effect} to conduct a mediation analysis in the presence of a potentially misclassified binary mediator variable. The current paper showcases novel extensions of the original methods. For example, we use regression parameter estimates from \textbf{COMBO} to construct a new misclassification-corrected predictive model. In addition, we demonstrate how functions from \textbf{COMBO} can allow users to conduct a receiver operating characteristic (ROC) analysis when outcome variables are misclassified, based on methods first proposed in Zawistowski et al. (2017)\cite{zawistowski2017corrected}. Using \textbf{COMMA}, we replicate an applied study that previously relied on known misclassification rates for the observed binary mediator variable.

Existing R packages for misclassified binary outcome variables have the same limitations of the methodologies they implement. For example, the \textbf{SAMBA} R package applies the methods proposed in Beesley and Mukherjee (2020)\cite{beesley2020statistical}, and thus relies on a perfect specificity assumption \citep{SAMBA}. To our knowledge, there are no other R software packages that enable association estimation in the presence of sequential and dependent noisy outcomes or in the presence of misclassified mediator variables. 

In Section \ref{COMBO-examples}, we provide two examples of the \textbf{COMBO} R package. First, we demonstrate \textbf{COMBO} methods for misclassified binary outcome variables using data from The Law School Admissions Council's 1998 National Bar Passage Study \citep{wightman1998lsac}. Using this data, we study predictors of bar exam passage and investigate whether passage rates of the bar exam depend on the examinee's race among individuals of identical skill. We also provide a demonstration of the methods for multiple misclassified sequential and dependent binary outcome variables using \textbf{COMBO} and pretrial detention data from Prince William County Virginia. In this example, we develop a misclassification-corrected risk prediction model for the study setting. In addition, we extend the ROC analysis approach from Zawistowski et al. (2017)\cite{zawistowski2017corrected} to compare our misclassification-corrected risk prediction model to an existing risk prediction algorithm. 

In Section \ref{comma-methods} we provide an overview of the methods for mediation analysis with a misclassified mediator variable. In Section \ref{comma-example}, we demonstrate \textbf{COMMA} through a replication study of the mediating effect of gestational hypertension on the relationship between maternal ethnicity and birthweight \citep{li_direct_2020}. Our replication allows for the direct estimation of gestational hypertension misdiagnosis rates, rather than relying on a sensitivity analysis approach \citep{li_direct_2020}. A conclusion is provided in Section \ref{discussion}.

\section[Methods]{Methods for misclassified binary outcome variables}\label{combo-methods}
In this section, we provide a brief overview of bias-correction methods for regression models with misclassified binary outcome variables. These methods are described in greater depth in Webb and Wells (2023)\cite{webb2023statistical} and Webb, Riley, and Wells (2023)\cite{webb2023assessment}.

\subsection[Misclassified binary outcome variables]{Misclassified binary outcome variables}\label{combo1-methods}
We are interested in the relationship between a matrix of predictors, $\boldsymbol{X}$, and a latent variable $Y$, which denotes true outcome status, taking values $j \in \{1,2\}$. The outcome $Y$ may be imperfectly measured and our observed proxy for $Y$ is denoted $Y^*$, which takes values $k \in \{1,2\}$, is a potentially misclassified version of $Y$. The matrix of predictors $\boldsymbol{Z}$ is related to sensitivity and specificity of our outcome measure. We assume that both $\boldsymbol{X}$ and $\boldsymbol{Z}$ are measured without error. We model these relationships as follows: 
\begin{equation}
\begin{aligned}
\label{eq:combo1-conceptual_framework_eq}
\text{True outcome mechanism: } &\; \logit\{ P(Y = j | \boldsymbol{X} ; \boldsymbol{\beta}) \} = \beta_{j0} + \boldsymbol{\beta_{jX} X}. \\
\text{Observation mechanisms: } &\; \logit\{ P(Y^* = k | Y = 1, \boldsymbol{Z} ; \boldsymbol{\gamma}) \} = \gamma_{k10} + \boldsymbol{\gamma_{k1Z} Z}, \\
                                &\; \logit\{ P(Y^* = k | Y = 2, \boldsymbol{Z} ; \boldsymbol{\gamma}) \} = \gamma_{k20} + \boldsymbol{\gamma_{k2Z} Z},
\end{aligned}
\end{equation}
Response probabilities for individual $i$'s latent outcome $Y$ and for individual $i$'s observed outcome $Y^*$, given $Y$, are expressed as: 
\begin{flalign}
\begin{aligned}
\label{eq:combo1-response_probabilities_eq}
P(Y_i = j | \boldsymbol{X_i} ; \boldsymbol{\beta}) = &\; \; \pi_{ij} = \frac{\text{exp}\{\beta_{j0} + \boldsymbol{\beta_{jX} X_i}\}}{1 + \text{exp}\{\beta_{j0} + \boldsymbol{\beta_{jX} X_i}\}}, \\
P(Y^*_i = k | Y_i = j, \boldsymbol{Z} ; \boldsymbol{\gamma}) = &\; \pi^*_{ikj} = \frac{\text{exp}\{\gamma_{kj0} + \boldsymbol{\gamma_{kjZ} Z_i}\}}{1 + \text{exp}\{\gamma_{kj0} + \boldsymbol{\gamma_{kjZ} Z_i}\}}.
\end{aligned}
\end{flalign}
where $\boldsymbol{\beta} = (\beta_{10}, \beta_{1x_1}, \dots, \beta_{1x_{p_x}})$ and $\boldsymbol{\gamma} = (\gamma_{110}, \gamma_{11{z_1}}, \dots, \gamma_{11{z_{p_z}}}, \gamma_{120}, \gamma_{12{z_1}}, \dots, \gamma_{12{z_{p_z}}})$. Average sensitivity, denoted $\pi^*_{11}$, is computed by taking the mean of $\pi^*_{ikj}$ when $k = 1$ and $j = 1$ across all $N$ subjects as follows, $\pi^*_{11} = \frac{1}{N}\sum_{i = 1}^N \pi^*_{i11}$. Similarly, we denote the average specificity as $\pi^*_{22} = \frac{1}{N}\sum_{i = 1}^N \pi^*_{i22}$. Note that these equations demonstrate that we can use the model forms in (\ref{eq:combo1-conceptual_framework_eq}) to express sensitivity and specificity as a function of the covariates $\boldsymbol{Z}$ in our dataset.

Based on this model and using the law of total probability, the probability of observing outcome category $k$ can be written as: 
\begin{equation}
\begin{aligned}
\label{eq:combo1-p_obs_Ystar}
P(Y^* = k | \boldsymbol{X}, \boldsymbol{Z}) = \sum_{j = 1}^2 P(Y^* = k | Y = j, \boldsymbol{Z} ; \boldsymbol{\gamma}) P(Y = j | \boldsymbol{X} ; \boldsymbol{\beta}) = \sum_{j = 1}^2 \pi^*_{kj} \pi_{j}.
\end{aligned}
\end{equation}
In (\ref{eq:combo1-p_obs_Ystar}), $\pi^*_{kj} = P(Y^* = k | Y = j, \boldsymbol{Z} ; \boldsymbol{\gamma})$ and $\pi_{j} = P(Y = j | \boldsymbol{X} ; \boldsymbol{\beta})$. For a single observation, $(Y_i, X_i, Z_i)$, the contribution to the likelihood is $\prod_{k = 1}^2 P(Y^*_i = k | \boldsymbol{X_i}, \boldsymbol{Z_i})^{y^*_{ik}}$ where $y^*_{ik} = \mathbbm{I}(Y^*_i = k)$ and $\mathbbm{I}(A)$ is the indicator of the set $A$.

The observed data log-likelihood for subjects $i = 1 \dots N$ is expressed as:
\begin{equation}
\begin{aligned}
\label{eq:combo1-obs-log-like}
\ell_{obs}(\boldsymbol{\beta}, \boldsymbol{\gamma}; \boldsymbol{X}, \boldsymbol{Z}) = \sum_{i = 1}^N \sum_{k = 1}^2 y^*_{ik} \text{log} \{ P(Y^*_i = k | \boldsymbol{X_i}, \boldsymbol{Z_i}) \} = \sum_{i = 1}^N \sum_{k = 1}^2 y^*_{ik} \text{log} \{ \sum_{j = 1}^2 \pi^*_{ikj} \pi_{ij} \}.
\end{aligned}
\end{equation}

Our goal is to estimate $(\boldsymbol{\beta}, \boldsymbol{\gamma})$ from (\ref{eq:combo1-conceptual_framework_eq}). The \textbf{COMBO} R package provides two estimation methods, an Expectation-Maximization (EM) algorithm and a Bayesian estimation scheme, which are described in Sections \ref{combo1-em} and \ref{combo1-mcmc}, respectively.

Note that direct optimization of the observed data log-likelihood in (\ref{eq:combo1-obs-log-like}) is another viable estimation strategy, but this method tends to be sensitive to starting parameter values. In addition, this approach tends to be unstable for large datasets and/or datasets with a moderate number of predictor variables. A discussion of these challenges is provided in Webb and Wells (2023)\cite{webb2023statistical}.

The estimates in the model described in (\ref{eq:combo1-obs-log-like}) suffers from the known problem of \textit{label switching} \citep{webb2023statistical}. Label switching occurs when likelihood functions for latent variable models are permutation invariant, meaning that multiple labelings of the latent variables result in identically-valued likelihood functions. Permutation invariance of the likelihood results in multiple parameter sets that maximize the likelihood, leading to challenges with estimation \citep{redner1984mixture}. The \textbf{COMBO} R package includes a built-in label switching correction procedure, which automatically selects the most plausible parameter set for the supplied data. The most plausible parameter set is operationalized as the parameter set which maximizes Youden's $J$ Statistic, a composite measure of the performance of a classifier \citep{BERRAR2019546}. Youden's $J$ Statistic is commonly used in misclassification models to deal with labeling degeneracy problems \citep{collins2014estimation, duan2021global, lamont2016regression, jones2010identifiability}. More details on the label switching correction procedure deployed in \textbf{COMBO} are provided in Webb and Wells (2023)\cite{webb2023statistical}.

\subsubsection{An EM algorithm for misclassified binary outcome models} \label{combo1-em}
For the Expectation-Maximization (EM) algorithm approach \citep{dempster1977maximum}, we begin with the complete data log-likelihood function:
\begin{equation}
    \begin{aligned}
    \label{eq:combo1-complete-log-like}
    \ell_{complete}(\boldsymbol{\beta}, \boldsymbol{\gamma}; \boldsymbol{X}, \boldsymbol{Z}) &= \sum_{i = 1}^N \Bigg[ \sum_{j = 1}^2 y_{ij} \text{log} \{ P(Y_i = j | \boldsymbol{X_i}) \} \\  &\qquad\phantom{a}  + \sum_{j = 1}^2 \sum_{k = 1}^2 y_{ij} y^*_{ik} \text{log} \{ P(Y^*_i = k | Y_i = j, \boldsymbol{Z_i}) \}\Bigg] & \\
    &= \sum_{i = 1}^N \Bigg[ \sum_{j = 1}^2 y_{ij} \text{log} \{ \pi_{ij} \} + \sum_{j = 1}^2 \sum_{k = 1}^2 y_{ij} y^*_{ik} \text{log} \{ \pi^*_{ikj} \}\Bigg],
    \end{aligned}
\raisetag{12pt}\end{equation}
where $y_{ij} = \mathbbm{I}(Y_i = j)$. Because this likelihood contains the latent binary outcome variable, $Y$, we cannot use it directly for maximization. 

For the expectation step or E-step of the algorithm, we compute the expected value of the latent variable, conditional on the observed data:
\begin{equation}
\begin{aligned}
\label{eq:combo1-e-step}
w_{ij} = P(Y_i = j | Y_i^*, \boldsymbol{X}, \boldsymbol{Z}) = \sum_{k = 1}^2 \frac{y^*_{ik} \pi^*_{ikj} \pi_{ij}}{\sum_{\ell = 1}^2 \pi^*_{i k \ell} \pi_{i \ell}}.  
\end{aligned}
\end{equation}
Since (\ref{eq:combo1-complete-log-like}) is linear in the latent variable, we can express the expected log-likelihood by replacing $y_{ij}$ in (\ref{eq:combo1-complete-log-like}) with $w_{ij}$, computed in (\ref{eq:combo1-e-step}):
\begin{equation}
\begin{aligned}
\label{eq:combo1-m-step}
Q = \sum_{i = 1}^N \Bigl[ \sum_{j = 1}^2 w_{ij} \text{log} \{ \pi_{ij} \} + \sum_{j = 1}^2 \sum_{k = 1}^2 w_{ij} y^*_{ik} \text{log} \{ \pi^*_{ikj} \}\Bigr].
\end{aligned}
\end{equation}
For the maximization step or M-step, we maximize the function in (\ref{eq:combo1-m-step}) with respect to $\boldsymbol{\beta}$ and $\boldsymbol{\gamma}$. Within \textbf{COMBO}, the M-step is performed separately for $\boldsymbol{\beta}$, $\boldsymbol{\gamma_{k1}}$, and $\boldsymbol{\gamma_{k2}}$, where $\boldsymbol{\gamma_{k p}}$ denotes the $\boldsymbol{\gamma}$ parameters where the second subscript is equal to $p \in \{1, 2\}$.

To estimate these parameter vectors separately, we split (\ref{eq:combo1-m-step}) into three equations as follows:
\begin{equation}
\begin{aligned}
\label{eq:combo1-q-split}
&Q_{\boldsymbol{\beta}} = \sum_{i = 1}^N \Bigl[ \sum_{j = 1}^2 w_{ij} \text{log} \{ \pi_{ij} \}\Bigr], \\
&Q_{\boldsymbol{\gamma_{k1}}} = \sum_{i = 1}^N \Bigl[\sum_{k = 1}^2 w_{i1} y^*_{ik} \text{log} \{ \pi^*_{ik1} \}\Bigr], 
Q_{\boldsymbol{\gamma_{k2}}} = \sum_{i = 1}^N \Bigl[\sum_{k = 1}^2 w_{i2} y^*_{ik} \text{log} \{ \pi^*_{ik2} \}\Bigr].
\end{aligned}
\end{equation}
To estimate $\boldsymbol{\beta}$, $Q_{\boldsymbol{\beta}}$ is fit as a logistic regression model where the outcome variable is the E-step term $w_{ij}$. To estimate $\boldsymbol{\gamma_{k1}}$, and $\boldsymbol{\gamma_{k2}}$, $Q_{\boldsymbol{\gamma_{k1}}}$ and $Q_{\boldsymbol{\gamma_{k2}}}$ are each, separately, fit as weighted logistic regression models where the outcome is $y^*_{ik}$ \citep{agresti2003categorical}. Final parameter estimates are obtained through a label switching correction procedure, which is automatically carried out on parameter estimates obtained through the EM algorithm in \textbf{COMBO} \citep{webb2023statistical}.

\subsubsection{Bayesian modeling for misclassified binary outcome models}\label{combo1-mcmc}
A Bayesian approach is also provided in \textbf{COMBO}. Our proposed binary outcome misclassification model is: $Y^*_{i} | \pi^*_{i} \sim Bernoulli(\pi^*_{i})$. Here, $\pi^*_{i} = \sum_{j = 1}^2 \pi^*_{i1j} \pi_{ij}$ as in (\ref{eq:combo1-p_obs_Ystar}). In \textbf{COMBO}, we provide functions that estimate this model using a Markov Chain Monte Carlo (MCMC) procedure. Users may select from multiple prior distribution settings, including Uniform, Normal, Double Exponential, or t prior distributions, with user-specified prior parameters. Recommendations for prior elicitation are provided in Webb and Wells (2023)\cite{webb2023statistical} Before summarizing the MCMC results, \textbf{COMBO} functions automatically apply a label switching correction procedure on \textit{each individual MCMC chain}. 

\subsection[Misclassified sequential and dependent binary outcome variables]{Misclassified sequential and dependent binary outcome variables}
In this section, we describe an extension to the model in \ref{combo1-methods} for a multistage framework. Once again, $Y = j$ denotes an observation's true outcome status, taking values $j \in \{1, 2\}$. Again, we are interested in the relationship between a matrix of predictors, $\boldsymbol{X}$ and the true, latent outcome $Y$. Instead of observing a single noisy proxy for $Y$, we observe two sequential and dependent imperfect measurements of $Y$, denoted $Y^{*(1)}$ and $Y^{*(2)}$. $Y^{*(1)}$ and $Y^{*(2)}$ are the first-stage and second-stage observed outcome variables, taking values in $k^{(1)} \in \{ 1, 2 \}$ and $k^{(2)} \in \{ 1, 2 \}$, respectively. $\boldsymbol{Z^{(1)}}$ denotes a predictor matrix for the misclassification of $Y^{*(1)}$ and $\boldsymbol{Z^{(2)}}$ denotes a predictor matrix for the misclassification of $Y^{*(2)}$. 

\begin{figure}[h!]
\begin{center}
\includegraphics[width = 460pt]{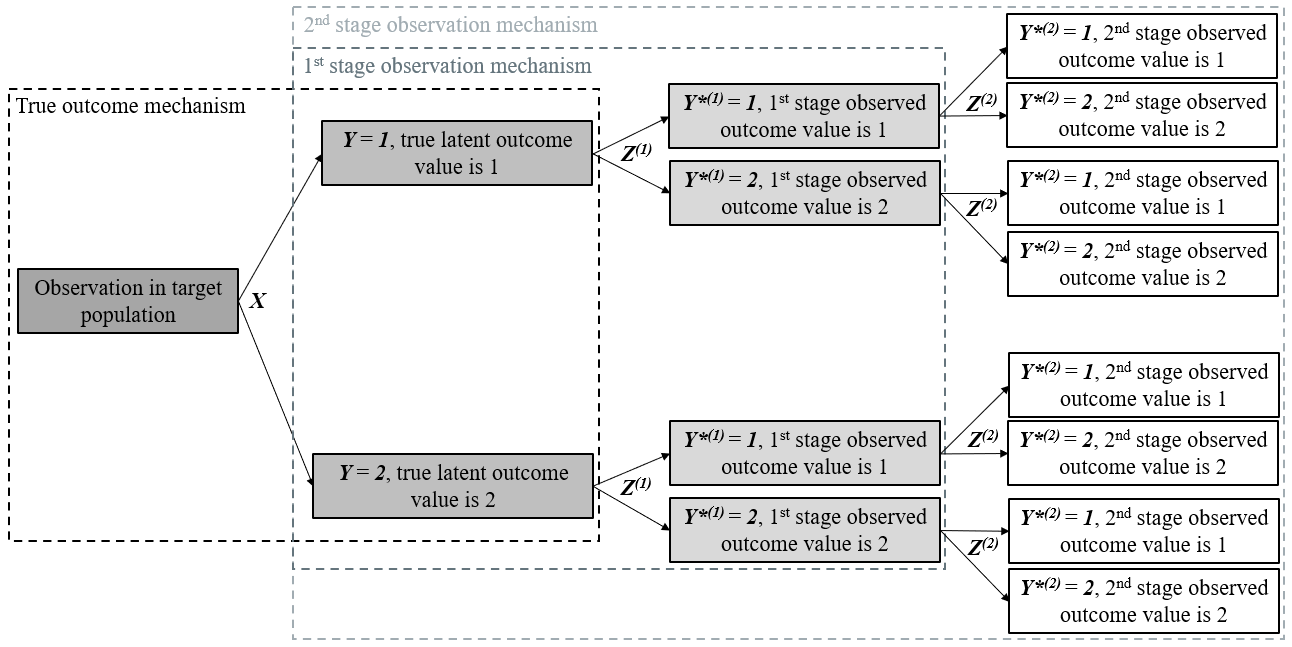}

\caption{Diagram of the assumed data structure for a two-stage misclassification model. Here, $\boldsymbol{X}$ is a set of predictors related to the true, latent outcome, $Y$. $\boldsymbol{Z^{(1)}}$ is a set of predictors related to the first-stage outcome $Y^{*(1)}$, conditional on the true (unobserved) outcome, $Y$. Similarly, $\boldsymbol{Z^{(2)}}$ is a set of predictors related to the second-stage outcome $Y^{*(2)}$, given the first-stage outcome $Y^{*(1)}$ and true (unobserved) outcome, $Y$. }\label{multistage_conceptual_framework_figure}
\end{center}
\end{figure}

Figure \ref{multistage_conceptual_framework_figure} presents the conceptual model, which can be mathematically expressed as follows:
\begin{equation}
\begin{aligned}
\label{eq:multistage_conceptual_framework_eq}
\text{True outcome mechanism: } &\; \text{logit}\{ P(Y = 1 | \boldsymbol{X} ; \boldsymbol{\beta}) \} = \beta_{0} + \boldsymbol{\beta}_{X} \boldsymbol{X}. \\
\\
\text{ First-stage observation mechanisms: } &\; \text{logit}\{ P(Y^{*(1)} = 1 | Y = 1, \boldsymbol{Z^{(1)}} ; \boldsymbol{\gamma}^{(1)}) \} \\ &\; \hspace{1.5em} = \gamma^{(1)}_{110} + \boldsymbol{\gamma^{(1)}_{11Z^{(1)}}} \boldsymbol{Z^{(1)}},\\
&\; \text{logit}\{ P(Y^{*(1)} = 1 | Y = 2, \boldsymbol{Z^{(1)}} ; \boldsymbol{\gamma}^{(1)}) \} \\ &\; \hspace{1.5em} = \gamma^{(1)}_{120} + \boldsymbol{\gamma^{(1)}_{12Z^{(1)}}} \boldsymbol{Z^{(1)}}. \\
\\
\text{Second-stage observation mechanisms: } &\; \text{logit}\{ P(Y^{*(2)} = 1 | Y^{*(1)} = 1, Y = 1, \boldsymbol{Z^{(2)}} ; \boldsymbol{\gamma}^{(2)}) \}  \\ &\; \hspace{1.5em} = \gamma^{(1)}_{1110} + \boldsymbol{\gamma^{(2)}_{111Z^{(2)}}} \boldsymbol{Z^{(2)}},\\
&\; \text{logit}\{ P(Y^{*(2)} = 1 | Y^{*(1)} = 2, Y = 1, \boldsymbol{Z^{(2)}} ; \boldsymbol{\gamma}^{(2)}) \} \\ &\; \hspace{1.5em} = \gamma^{(1)}_{1210} + \boldsymbol{\gamma^{(2)}_{121Z^{(2)}}} \boldsymbol{Z^{(2)}},\\
&\; \text{logit}\{ P(Y^{*(2)} = 1 | Y^{*(1)} = 1, Y = 2, \boldsymbol{Z^{(2)}} ; \boldsymbol{\gamma}^{(2)}) \} \\ &\; \hspace{1.5em} = \gamma^{(1)}_{1120} + \boldsymbol{\gamma^{(2)}_{112Z^{(2)}}} \boldsymbol{Z^{(2)}},\\
&\; \text{logit}\{ P(Y^{*(2)} = 1 | Y^{*(1)} = 2, Y = 2, \boldsymbol{Z^{(2)}} ; \boldsymbol{\gamma}^{(2)}) \} \\ &\; \hspace{1.5em} = \gamma^{(1)}_{1220} + \boldsymbol{\gamma^{(2)}_{122Z^{(2)}}} \boldsymbol{Z^{(2)}}.
\end{aligned}
\end{equation}

Such a model structure is plausible in several decision-making contexts in healthcare and criminal justice systems. For example, $Y$ may be an individual's true disease status or pretrial failure status. $Y^{*(1)}$ may represent a fallible test result or algorithmic recommendation. Using this first-stage result, a final fallible decision-maker $Y^{*(2)}$, like a doctor or a judicial officer, may reach a decision by either overriding or agreeing with $Y^{*(1)}$. Both of the potentially misclassified outcomes, $Y^{*(1)}$ and $Y^{*(2)}$, may be related to predictors $\boldsymbol{Z^{(1)}}$ and $\boldsymbol{Z^{(2)}}$, conditional on earlier-stage outcomes.

The probability of true, latent outcome category $j$ for individual $i$ is denoted $\pi_{ij}$. The probability of first-stage outcome $k$, conditional on true outcome status $j$ is denoted $\pi^{*(1)}_{i k j}$. The probability of second-stage outcome $\ell$, conditional on first-stage outcome $k$ and true outcome status $j$ is denoted $\pi^{*(2)}_{i \ell k j}$. Using (\ref{eq:multistage_conceptual_framework_eq}), we write these response probabilities for all $i = 1, \dots, N$ in the sample as follows:
\begin{flalign}
\begin{aligned}
\label{eq:multistage_response_probabilities_eq}
P(Y_i = j | \boldsymbol{X_i} ; \boldsymbol{\beta}) = &\; \; \pi_{ij} = \frac{\text{exp}\{\beta_{j0} + \boldsymbol{\beta_{jX} X_i}\}}{1 + \text{exp}\{\beta_{j0} + \boldsymbol{\beta_{jX} X_i}\}}, \\
P(Y^{*(1)}_i = k | Y_i = j, \boldsymbol{Z^{(1)}} ; \boldsymbol{\gamma^{(1)}}) = &\; \pi^{*(1)}_{i k j} = \frac{\text{exp}\{\gamma^{(1)}_{kj0} + \boldsymbol{\gamma^{(1)}_{kjZ} Z^{(1)}_i}\}}{1 + \text{exp}\{\gamma^{(1)}_{kj0} + \boldsymbol{\gamma^{(1)}_{kjZ} Z^{(1)}_i}\}}, \\
P(Y^{*(2)}_i = \ell | Y^{*(1)}_i = k, Y_i = j, \boldsymbol{Z^{(2)}} ; \boldsymbol{\gamma^{(2)}}) = &\; \pi^{*(2)}_{i \ell k j} = \frac{\text{exp}\{\gamma^{(2)}_{\ell kj0} + \boldsymbol{\gamma^{(2)}_{\ell kjZ} Z^{(2)}_i}\}}{1 + \text{exp}\{\gamma^{(2)}_{\ell kj0} + \boldsymbol{\gamma^{(2)}_{\ell kjZ} Z^{(2)}_i}\}}.
\end{aligned}
\end{flalign}

First-stage outcome sensitivity and specificity are computed as $\frac{1}{N}\sum_{i = 1}^N \pi^{*(1)}_{i11} = \pi^{*(1)}_{11}$ and $\frac{1}{N}\sum_{i = 1}^N \pi^{*(1)}_{i22} = \pi^{*(1)}_{22}$, respectively. By marginalizing over the first-stage outcome, we can also compute the sensitivity and specificity of the second-stage outcome as $\frac{1}{N}\sum_{i = 1}^N \sum_{k = 1}^2 \pi^{*(2)}_{i 1 k 1} \pi^{*(1)}_{ik1} = \pi^{*(2)}_{i11}$ and $\frac{1}{N}\sum_{i = 1}^N \sum_{k = 1}^2 \pi^{*(2)}_{i 2 k 2} \pi^{*(1)}_{ik2} = \pi^{*(2)}_{i22}$, respectively.

The probability of observing first-stage outcome $k$ and second-stage outcome $\ell$ jointly is:
\begin{equation}
\begin{aligned}
\label{eq:multistage_p_obs_Ystar}
P(Y^{*(1)} = k, Y^{*(2)} = \ell | \boldsymbol{X}, \boldsymbol{Z^{(1)}}, \boldsymbol{Z^{(2)}})   &\;
= \sum_{j = 1}^2 \Bigl( P(Y^{*(2)} = \ell | Y^{*(1)} = k, Y = j, \boldsymbol{Z^{(2)}}; \boldsymbol{\gamma^{(2)}}) \\
&\qquad\phantom{a} \times P(Y^{*(1)} = k | Y = j, \boldsymbol{Z^{(1)}}; \boldsymbol{\gamma^{(1)}})  \times P(Y = j | \boldsymbol{X} ; \boldsymbol{\beta}) \Bigr)
\\ & = \sum_{j = 1}^2 \Bigl( \pi^{*(2)}_{\ell kj} \times \pi^{*(1)}_{kj} \times \pi_{j} \Bigr),
\end{aligned}
\end{equation}
where $\pi^{*(2)}_{\ell kj} = P(Y^{*(2)} = \ell | Y^{*(1)} = k, Y = j, \boldsymbol{Z^{(2)}}; \boldsymbol{\gamma^{(2)}})$, $\pi^{*(1)}_{kj} = P(Y^{*(1)} = k | Y = j, \boldsymbol{Z^{(1)}}; \boldsymbol{\gamma^{(1)}})$ and $\pi_{j} = P(Y = j | \boldsymbol{X} ; \boldsymbol{\beta})$. 

For an individual observation $i$, the contribution to the likelihood is $\prod_{\k = 1}^2 \prod_{\ell = 1}^2 P(Y_i^{*(1)} = k, Y_i^{*(2)} = \ell | \boldsymbol{X_i}, \boldsymbol{Z_i^{(1)}}, \boldsymbol{Z_i^{(2)}})^{(y^{*(1)}_{ik} \times y^{*(2)}_{i\ell})}$ where $y^{*(1)}_{ik} = \mathbbm{I}(Y^{*(1)}_i = k)$, $y^{*(2)}_{i\ell} = \mathbbm{I}(Y^{*(2)}_i = \ell)$, and $\mathbbm{I}(A)$ is the indicator of the set $A$.

For the two-stage model, the observed data log-likelihood for observations $i = 1, \dots, N$ is:
\begin{equation}
\begin{aligned}
\label{eq:combo2-obs-log-like}
&\ell_{obs}(\boldsymbol{\beta}, \boldsymbol{\gamma^{(1)}}, \boldsymbol{\gamma^{(2)}}; \boldsymbol{X}, \boldsymbol{Z^{(1)}}, \boldsymbol{Z^{(2)}}) = \\
&\qquad\phantom{a} \sum_{i = 1}^N \sum_{k = 1}^2 \sum_{\ell = 1}^2 (y^{*(1)}_{ik} \times y^{*(2)}_{i\ell}) \times \text{log} \{ P(Y_i^{*(1)} = k, Y_i^{*(2)} = \ell | \boldsymbol{X_i}, \boldsymbol{Z_i^{(1)}}, \boldsymbol{Z_i^{(2)}}) \}
\\&\qquad\phantom{a} = \sum_{i = 1}^N \sum_{k = 1}^2 \sum_{\ell = 1}^2 (y^{*(1)}_{ik} \times y^{*(2)}_{i\ell}) \times \text{log} \{ \sum_{j = 1}^2 \pi^{*(2)}_{i\ell kj} \times \pi^{*(1)}_{ikj} \times \pi_{ij} \}.
\end{aligned}
\end{equation}

We aim to estimate $(\boldsymbol{\beta}, \boldsymbol{\gamma^{(1)}}, \boldsymbol{\gamma^{(2)}})$ from (\ref{eq:multistage_conceptual_framework_eq}). The \textbf{COMBO} R Package includes two estimation methods, an EM algorithm and a Bayesian estimation procedure, which are described in Sections \ref{combo2-em} and \ref{combo2-mcmc}, respectively. As in (\ref{eq:combo1-obs-log-like}), the structure of the model described in (\ref{eq:combo2-obs-log-like}) requires a label switching correction. The \textbf{COMBO} R Package automatically applies the label switching correction described in detail in Webb, Riley, and Wells (2023)\cite{webb2023assessment}.

In this section, we presented a model with two outcome stages, but the model can be extended to an arbitrary number of outcome stages \citep{webb2023assessment}. The \textbf{COMBO} R Package only supports two-stage outcome models.

\subsubsection{An EM algorithm for two-stage misclassified binary outcome models}\label{combo2-em}
For the two-stage model, we first present an Expectation-Maximization (EM) algorithm to jointly estimate $(\boldsymbol{\beta}, \boldsymbol{\gamma^{(1)}}, \boldsymbol{\gamma^{(2)}})$ \citep{dempster1977maximum}. We begin with the complete data log-likelihood function: 
\begin{equation}
\begin{aligned}
\label{eq:multistage-complete-log-like}
&\ell_{\text {complete }}(\boldsymbol{\beta}, \boldsymbol{\gamma^{(1)}}, \boldsymbol{\gamma^{(2)}} ; \boldsymbol{X}, \boldsymbol{Z^{(1)}}, \boldsymbol{Z^{(2)}})= \\
&\qquad\phantom{a} \sum_{i=1}^N \Bigl[\sum_{j=1}^2 y_{i j} \log \{P(Y_i=j \mid \boldsymbol{X}_{\boldsymbol{i}})\}\\ &\qquad\phantom{a} +\sum_{j=1}^2 \sum_{k=1}^2 y_{i j} y_{i k}^{*(1)} \log \{P(Y_i^{*(1)}=k \mid Y_i=j, \boldsymbol{Z_i^{(1)}})\} \\ &\qquad\phantom{a} +\sum_{j=1}^2 \sum_{k=1}^2 \sum_{\ell=1}^2 y_{i j} y_{i k}^{*(1)} y_{i l}^{*(2)} \log \{P(Y_i^{*(2)}=\ell \mid Y_i^{*(1)}=k, Y_i=j, \boldsymbol{Z_i^{(2)}}\}\Bigr],
\end{aligned}
\end{equation}
where $y_{ij} = \mathbbm{I}(Y_i = j)$. Since the true outcome $Y$ is a latent variable, we cannot use (\ref{eq:multistage-complete-log-like}) directly for maximization. Instead, we develop an EM algorithm with the E-step and M-step provided below. 

The E-step for a two-stage misclassified binary outcome model is:
\begin{equation}
\begin{aligned}
\label{multistage-e-step}
&w_{i j}= P(Y_i=j \mid Y_i^{*(2)}, Y_i^{*(1)}, \boldsymbol{X_i},  \boldsymbol{Z_i^{(1)}}, \boldsymbol{Z_i^{(2)}})= \\ &\sum_{k=1}^2 \sum_{\ell=2}^2 \frac{y_{i k}^{*(1)} y_{i \ell}^{*(2)} P(Y_i=j \mid \boldsymbol{X_i}) P(Y_i^{*(1)}=k \mid Y_i=j, \boldsymbol{Z_i^{(1)}}) P(Y_i^{*(2)}=\ell \mid Y_i^{*(1)}=k, Y_i=j, \boldsymbol{Z_i^{(2)}})}{\sum_{b=2}^2 P(Y_i=b \mid \boldsymbol{X_i}) P(Y_i^{*(1)}=k \mid Y_i=b,\boldsymbol{Z_i^{(1)}}) P(Y_i^{*(2)}=\ell \mid Y_i^{*(1)}=k, Y_i=b, \boldsymbol{Z_i^{(2)}})}.
\end{aligned}
\end{equation}

Because (\ref{eq:multistage-complete-log-like}) is linear in the latent variable, we obtain the function for the M-step by replacing every instance of $y_{ij}$ in (\ref{eq:multistage-complete-log-like}) with its expectation, $w_{ij}$, in (\ref{multistage-e-step}). The function to be maximized with respect to $(\boldsymbol{\beta}, \boldsymbol{\gamma^{(1)}}, \boldsymbol{\gamma^{(2)}})$ in the M-step, $Q$, is defined as:
\begin{equation}
\begin{aligned}
\label{eq:multistage-m-step}
Q = \sum_{i=1}^N & \Bigl[\sum_{j=1}^2 w_{i j} \log \{P(Y_i=j \mid \boldsymbol{X}_{\boldsymbol{i}})\} +\sum_{j=1}^2 \sum_{k=1}^2 w_{i j} y_{i k}^{*(1)} \log \{P(Y_i^{*(1)}=k \mid Y_i=j, \boldsymbol{Z_i^{(1)}})\} \\ & +\sum_{j=1}^2 \sum_{k=1}^2 \sum_{\ell=1}^2 w_{i j} y_{i k}^{*(1)} y_{i l}^{*(2)} \log \{P(Y_i^{*(2)}=\ell \mid Y_i^{*(1)}=k, Y_i=j, \boldsymbol{Z_i^{(2)}}\}\Bigr].
\end{aligned}
\end{equation}
$Q$ can also be written separately for the parameter vectors $\boldsymbol{\beta}, \boldsymbol{\gamma_{k1}^{(1)}}, \boldsymbol{\gamma_{k2}^{(1)}}, \boldsymbol{\gamma_{\ell 11}^{(2)}}, \boldsymbol{\gamma_{\ell 21}^{(2)}}, \boldsymbol{\gamma_{\ell 12}^{(2)}},$ and $\boldsymbol{\gamma_{\ell 22}^{(2)}}$. Here, $\boldsymbol{\gamma_{k1}^{(1)}}$ denotes $\boldsymbol{\gamma^{(1)}}$ parameters where the second subscript is set to $p \in \{1,2\}$ and where $\boldsymbol{\gamma_{\ell q p}^{(2)}}$ denotes $\boldsymbol{\gamma^{(2)}}$ parameters where the second and third subscripts are set to $q \in \{1,2\}$ and $p \in \{1,2\}$, respectively.
\begin{equation}
\begin{aligned}
\label{eq:combo2-q-split}
&Q_{\boldsymbol{\beta}} = \sum_{i = 1}^N \Bigl[ \sum_{j = 1}^2 w_{ij} \text{log} \{ \pi_{ij} \}\Bigr], \\ &Q_{\boldsymbol{\gamma^{(1)}_{k1}}} = \sum_{i = 1}^N \Bigl[\sum_{k = 1}^2 w_{i1} y^{*(1)}_{ik} \text{log} \{ \pi^{*(1)}_{ik1} \}\Bigr],
Q_{\boldsymbol{\gamma^{(1)}_{k2}}} = \sum_{i = 1}^N \Bigl[\sum_{k = 1}^2 w_{i2} y^{*(1)}_{ik} \text{log} \{ \pi^{*(1)}_{ik2} \}\Bigr],\\
&Q_{\boldsymbol{\gamma^{(2)}_{\ell 11 }}} = \sum_{i = 1}^N \Bigl[\sum_{\ell = 1}^2 w_{i1} y^{*(1)}_{i1} y^{*(2)}_{i\ell} \text{log} \{ \pi^{*(2)}_{i \ell 11} \}\Bigr], Q_{\boldsymbol{\gamma^{(2)}_{\ell 21 }}} = \sum_{i = 1}^N \Bigl[\sum_{\ell = 1}^2 w_{i1} y^{*(1)}_{i2} y^{*(2)}_{i\ell} \text{log} \{ \pi^{*(2)}_{i \ell 21} \}\Bigr], \\
&Q_{\boldsymbol{\gamma^{(2)}_{\ell 12 }}} = \sum_{i = 1}^N \Bigl[\sum_{\ell = 1}^2 w_{i2} y^{*(1)}_{i1} y^{*(2)}_{i\ell} \text{log} \{ \pi^{*(2)}_{i \ell 12} \}\Bigr], Q_{\boldsymbol{\gamma^{(2)}_{\ell 22 }}} = \sum_{i = 1}^N \Bigl[\sum_{\ell = 1}^2 w_{i2} y^{*(1)}_{i2} y^{*(2)}_{i\ell} \text{log} \{ \pi^{*(2)}_{i \ell 22} \}\Bigr].
\end{aligned}
\end{equation}

Within \textbf{COMBO}, $\boldsymbol{\beta}$ in (\ref{eq:combo2-q-split}) is estimated by fitting $Q_{\boldsymbol{\beta}}$ as a logistic regression model with outcome variable $w_{ij}$. $Q_{\boldsymbol{\gamma^{(1)}_{kp}}}$ for $p \in \{1,2\}$ are separately fit as weighted logistic regression models with outcome variable $y^{*(1)}_{ik}$. Similarly, $Q_{\boldsymbol{\gamma^{(2)}_{\ell qp}}}$ for $p \in \{1,2\}$ and $q \in \{1,2\}$ are separately fit as weighted logistic regression models with outcome variable $y^{*(1, 2)}_{i\ell q} = y^{*(1)}_{iq} \times y^{*(2)}_{i\ell}$ \citep{agresti2003categorical}. A label switching correction procedure is required to obtain final prameter estimates \citep{webb2023assessment}. In the \textbf{COMBO} R Package, this correction is deployed automatically on parameter estimtes obtained from this EM algorithm.

\subsubsection{Bayesian modeling for two-stage misclassified binary outcome models}\label{combo2-mcmc}
The \textbf{COMBO} R Package also provides a Bayesian estimation procedure to fit the model in (\ref{eq:multistage_conceptual_framework_eq}). Our proposed multistage misclassification model is defined for each of the stages in the model: $Y^{* (1)}_i | \pi^{* (1)}_i \sim Bernoulli ( \pi^{* (1)}_i)$ and $Y^{* (2)}_i | \pi^{* (2)}_i \sim Bernoulli ( \pi^{* (2)}_i)$. We define the response probabilities for both model stages as follows:
\begin{equation}
\begin{aligned}
\label{eq:multistage_p_obs_Ystar1}
\pi^{* (1)}_i &\;= P(Y_i^{*(1)} = 1 | \boldsymbol{X_i}, \boldsymbol{Z_i^{(1)}}) 
= \sum_{j = 1}^2 \Bigl[ P(Y_i^{*(1)} = 1 | Y_i = j, \boldsymbol{Z_i^{(1)}}; \boldsymbol{\gamma^{(1)}})  \times P(Y_i = j | \boldsymbol{X_i} ; \boldsymbol{\beta}) \Bigr]
\\ & = \sum_{j = 1}^2 \Bigl( \pi^{*(1)}_{i1j} \times \pi_{ij} \Bigr), 
\end{aligned}
\end{equation}
\begin{equation}
\begin{aligned}
\label{eq:multistage_p_obs_Ystar2}
\pi^{* (2)}_i &\;= P(Y^{*(2)} = 1| \boldsymbol{X_i}, \boldsymbol{_iZ^{(1)}}, \boldsymbol{Z_i^{(2)}}) 
& \\
& = \sum_{j = 1}^2 \sum_{k = 1}^2 \Bigl[ P(Y^{*(2)} = 1 | Y^{*(1)} = k, Y = j, \boldsymbol{Z_i^{(2)}}; \boldsymbol{\gamma^{(2)}}) \\
&\qquad\phantom{a} \qquad\phantom{a} \times P(Y_i^{*(1)} = k | Y_i = j, \boldsymbol{Z_i^{(1)}}; \boldsymbol{\gamma^{(1)}})  \times P(Y_i = j | \boldsymbol{X_i} ; \boldsymbol{\beta}) \Bigr]
\\ & = \sum_{j = 1}^2 \sum_{k = 1}^2 \Bigl( \pi^{*(2)}_{i 1 kj} \times \pi^{*(1)}_{ikj} \times \pi_{ij} \Bigr).
\end{aligned}
\end{equation}
We employ a Markov Chain Monte Carlo (MCMC) procedure to estimate $(\boldsymbol{\beta}, \boldsymbol{\gamma^{(1)}}, \boldsymbol{\gamma^{(2)}})$. Strategies for selecting prior distributions are discussed in Webb, Riley, and Wells (2023)\cite{webb2023assessment}. In the R Package, \textbf{COMBO}, users can specify prior parameters for either Uniform, Normal, Double Exponential, or $t-$ prior distributions. Within each MCMC chain, a necessary label switching correction procedure is automatically applied by the \textbf{COMBO} MCMC functions. Details on the correction are provided in Webb, Riley, and Wells (2023)\cite{webb2023assessment}.

\section[The COMBO package]{The \textbf{COMBO} package}\label{COMBO-examples}

\subsection[Installation and basic usage]{Installation and basic usage}
The \textbf{COMBO} package is available on CRAN, and information on the package can be found at \href{https://cran.r-project.org/web/packages/COMBO/index.html}{https://cran.r-project.org/web/packages/COMBO/index.html}. To install and load \textbf{COMBO} in \texttt{R}, run the following code:
\begin{lstlisting}
R> install.packages("COMBO")
R> library(COMBO)
\end{lstlisting}

\subsection{Applying COMBO to study bias in the bar exam}
In this section, we present a case study to examine predictors of successful bar exam passage (an examination that a lawyer must pass in order to legally practice in a given jurisdiction), as well as whether the bar exam is biased against Black students. The data for this case study was originally collected in 1998 for The Law School Admissions Council's (LSAC) National Bar Passage Study \citep{wightman1998lsac}. It is available in \textbf{COMBO} as the \texttt{LSAC\_data} and can be loaded as follows:
\begin{lstlisting}
R> data("LSAC_data")
\end{lstlisting}

First, we prepare the data for analysis by removing all missing values for variables of interest using the functions in the \textbf{dplyr} R package \citep{dplyr}. In addition, we limit the data to only include students with the \texttt{race1} variable recorded as \texttt{"white"} or \texttt{"black"}. We also create a centered and scaled version of the \texttt{lsat} score variable and add a subject number for each individual in the dataset. 

\begin{lstlisting}
R> #install.packages("dplyr")
R> library(dplyr)
R> LSAC_data_clean <- LSAC_data %>%
R>   filter(race1 %in% c("black", "white")) %>%
R>   filter(!is.na(bar_passed)) %>%
R>   filter(!is.na(lsat)) %>%
R>   filter(!is.na(zgpa)) %>%
R>   mutate(lsat_c = (lsat - mean(lsat)) / sd(lsat) %>%
R>   mutate(Subject = 1:n())
\end{lstlisting}

\subsubsection{Parameter estimation}
We wish to fit the following true outcome mechanism and observation mechanism:
\begin{equation}
\begin{aligned}
\label{eq:bar-conceptual_framework_eq}
\text{True outcome mechanism: } &\; \logit\{ P(Y = 1 | \boldsymbol{X} ; \boldsymbol{\beta}) \} = \beta_{1} + \beta_{2} \textsf{lsat\_c} + \beta_{3} \textsf{zgpa}. \\
\text{Observation mechanisms: } &\; \logit\{ P(\textsf{bar\_passed} = 1| Y = 1, \boldsymbol{Z} ; \boldsymbol{\gamma}) \} = \gamma_{11} + \gamma_{21} \textsf{race1}, \\
                                &\; \logit\{ P(\textsf{bar\_passed} = 1 | Y = 2, \boldsymbol{Z} ; \boldsymbol{\gamma}) \} = \gamma_{12} + \gamma_{22} \textsf{race1}.
\end{aligned}
\end{equation}
Here, $Y$ represents a latent indicator variable for ``admission to the bar". We want to model the relationship $Y$ and students' standardized LSAT scores ($\textsf{lsat\_c}$) and law school GPA ($\textsf{zgpa}$) in the \textit{true outcome mechanism}. We view the variable $\textsf{bar\_passed}$ as a noisy, potentially misclassified proxy for $Y$. In the \textit{observation mechanisms}, we model classification of $\textsf{bar\_passed}$, conditional on true and latent bar admission status, as a function of students' race ($\textsf{race1}$).

Before fitting this model using \textbf{COMBO}, we must convert the observed outcome variable from a logical variable to an indicator variable. In addition, we convert the \texttt{z} variable from a character variable to an indicator variable. Finally, we prepare the predictor matrix for the true outcome mechanism.
\begin{lstlisting}
R> bar_ystar <- matrix(ifelse(LSAC_data_clean$bar_passed == TRUE,
R>                     1, 2), ncol = 1)
R> bar_z <- matrix(ifelse(LSAC_data_clean$race1 == "white",
R>                        0, 1), ncol = 1)
R> bar_x <- matrix(c(LSAC_data_clean$lsat_c,
R>                   LSAC_data_clean$zgpa),
R>                 ncol = 2, byrow = FALSE)
\end{lstlisting}
To obtain starting values for the EM algorithm, we fit a naive logistic regression model for the observed outcome \citep{webb2023statistical}.
\begin{lstlisting}
R> bar_ystar_01 <- ifelse(LSAC_data_clean$bar_passed == TRUE, 1, 0)
R> bar_beta_glm <- glm(bar_ystar_01 ~ bar_x[,1] + bar_x[,2],
R>                     family = "binomial"(link = "logit"))
R> bar_beta_start <- matrix(unname(coef(bar_beta_glm)), ncol = 1)
\end{lstlisting}
Next, we use the \texttt{COMBO\_EM} function to obtain parameter estimates from our proposed EM algorithm. 
\begin{lstlisting}
R> bar_EM <- COMBO_EM(Ystar = bar_ystar[,1], 
R>                    x_matrix = bar_x, 
R>                    z_matrix = bar_z, 
R>                    beta_start = bar_beta_start,
R>                    gamma_start = matrix(rep(1, 4), ncol = 2))
\end{lstlisting}

In addition to our proposed EM algorithm, \texttt{COMBO\_EM} reports parameter estimates and standard errors (SE) from three comparison models: (1) \texttt{SAMBA}, which assumes perfect specificity in the observation mechanism \citep{SAMBA, beesley2020statistical}, (2) \texttt{PSens}, which assumes perfect sensitivity in the observation mechanism, and (3) \texttt{naive}, which assumes no misclassification in the observed outcome variable. Note that the \texttt{naive} method is equivalent to the logistic regression model used to obtain starting $\boldsymbol{\beta}$ parameter values for the EM algorithm. For each estimation method, the output from \texttt{COMBO\_EM} displays the estimate and standard error (SE) estimate for each parameter in the model in a data frame. In addition, the \texttt{Convergence} output column indicates whether or not convergence was reached for the EM algorithm and naive approaches. 

\begin{lstlisting}
R> bar_EM
R>        Parameter    Estimates           SE Convergence
R> 1          beta1   3.48220018 3.655382e-02        TRUE
R> 2          beta2   0.76200084 3.695353e-02        TRUE
R> 3          beta3   1.67913508 3.636389e-02        TRUE
R> 4        gamma11   7.99699335 2.871584e-01        TRUE
R> 5        gamma21  -2.22008994 4.855689e-01        TRUE
R> 6        gamma12   0.09942726 1.005696e-01        TRUE
R> 7        gamma22  -0.34736524 2.607421e-01        TRUE
R> 8    SAMBA_beta1   4.07734229 8.636230e-02          NA
R> 9    SAMBA_beta2   0.63480753 3.830245e-02          NA
R> 10   SAMBA_beta3   1.36816697 5.921907e-02          NA
R> 11 SAMBA_gamma11  20.43966759 5.088680e+05          NA
R> 12 SAMBA_gamma21 -16.48407514 5.088680e+05          NA
R> 13   PSens_beta1   3.45312894 2.718754e-01          NA
R> 14   PSens_beta2   0.74708681 8.111720e-02          NA
R> 15   PSens_beta3   1.63368299 1.512236e-01          NA
R> 16 PSens_gamma12   0.05999081 2.274230e-02          NA
R> 17 PSens_gamma22  -0.37100032 8.692969e-02          NA
R> 18   naive_beta1   4.05547394 6.436355e-02        TRUE
R> 19   naive_beta2   0.65462338 3.599356e-02        TRUE
R> 20   naive_beta3   1.35694859 4.967585e-02        TRUE
\end{lstlisting}

We can also fit the model in (\ref{eq:bar-conceptual_framework_eq}) with a Bayesian approach in \textbf{COMBO} using the \texttt{COMBO\_MCMC} function. First, we define the prior distributions for our model parameters. 
\begin{lstlisting}
R> prior_distribution <- "normal"
R> normal_mu_beta <- matrix(c(0, 0, 0, NA, NA, NA),
R>                           nrow = 2, byrow = TRUE)
R> normal_sd_beta <- matrix(c(10, 10, 10, NA, NA, NA),
R>                           nrow = 2, byrow = TRUE)
R> normal_mu_gamma <- array(data = c(0, NA, 0, NA, 0, NA, 0, NA),
R>                           dim = c(2,2,2))
R> normal_sd_gamma <- array(data = c(10, NA, 10, NA, 10, NA, 10, NA),
R>                           dim = c(2,2,2))
R> beta_prior_parameters <- list(mu = normal_mu_beta,
R>                               sigma = normal_sd_beta)
R> gamma_prior_parameters <- list(lower = normal_mu_gamma,
R>                                upper = normal_sd_gamma)
\end{lstlisting}
Next, we use the \texttt{COMBO\_MCMC} function to obtain parameter estimates from our proposed MCMC approach. Our demonstration code uses a relatively small number of chains and samples to ensure reasonable run time. 
\begin{lstlisting}
R> bar_MCMC <- COMBO_MCMC(Ystar = bar_ystar[,1], 
R>                        x_matrix = bar_x,
R>                        z_matrix = bar_z,
R>                        prior = prior_distribution,
R>                        beta_prior_parameters = beta_prior_parameters,
R>                        gamma_prior_parameters = gamma_prior_parameters,
R>                        number_MCMC_chains = 2,
R>                        MCMC_sample = 1000,
R>                        burn_in = 500)
\end{lstlisting}

Our \texttt{bar\_MCMC} output is a list containing the posterior samples and posterior means for both the binary outcome misclassification model and a naive logistic regression model. 
\begin{lstlisting}
R> bar_MCMC$posterior_means_df
R> # A tibble: 7 x 3
R>   parameter_name posterior_mean posterior_median
R>   <fct>                   <dbl>            <dbl>
R> 1 beta[1,1]              3.27             3.27  
R> 2 beta[1,2]              0.798            0.799
R> 3 beta[1,3]              1.71             1.71  
R> 4 gamma[1,1,1]           4.98             4.97  
R> 5 gamma[1,2,1]           0.518            0.519 
R> 6 gamma[1,1,2]           0.0647           0.0453
R> 7 gamma[1,2,2]          -0.487           -0.492 
\end{lstlisting}

\begin{lstlisting}
R> bar_MCMC$naive_posterior_means_df
R> # A tibble: 3 x 3
R>   parameter_name  posterior_mean posterior_median
R>   <chr>                    <dbl>            <dbl>
R> 1 naive_beta[1,1]          3.88             3.89 
R> 2 naive_beta[1,2]          0.624            0.623
R> 3 naive_beta[1,3]          1.26             1.26
\end{lstlisting}

We can use our $\boldsymbol{\gamma}$ parameter estimates and the \textbf{COMBO} \texttt{misclassification\_prob} function to compute average classification rate estimates by student race in our data. First, we compute these probabilities using $\boldsymbol{\gamma}$ parameter estimates from our proposed EM algorithm. 
\begin{lstlisting}
R> bar_misclass_probs_EM <- misclassification_prob(
R>                              matrix(bar_EM$Estimates[4:7],
R>                                     ncol = 2), 
R>                              bar_z)
R> bar_misclass_probs_EM %>%
R>   left_join(bar_data_clean %>% select(Subject, race1),
R>             by = "Subject") %>%
R>   group_by(Y, Ystar, race1) %>%
R>   summarise(mean_pistar = mean(Probability)) %>%
R>   ungroup()
R> # A tibble: 8 x 4
R>       Y Ystar race1 mean_pistar
R>   <int> <int> <chr>       <dbl>
R> 1     1     1 black       0.997
R> 2     1     1 white           1
R> 3     1     2 black       0.003
R> 4     1     2 white           0
R> 5     2     1 black       0.438
R> 6     2     1 white       0.525
R> 7     2     2 black       0.562
R> 8     2     2 white       0.475
\end{lstlisting}
In the resulting table, \texttt{Y} represents the value of the latent construct ``admission to the bar'', where the value \texttt{1} indicates that the student \textit{should have been} admitted to the bar and the value \texttt{2} indicates that the student \textit{should not have been} admitted to the bar. The \texttt{Ystar} column corresponds to observed bar exam passage from the data, where \texttt{1} indicates passing and \texttt{2} indicates failing. The \texttt{mean\_pistar} column denotes the average probability of being classified with value \texttt{Ystar} given true value \texttt{Y} among students of a given race. We can compute similar probabilities for all presented methods. Results are summarized in Table \ref{bar-probability-results-table}.   

\begin{table}[htbt]
\centering
\caption{Estimated sensitivity and specificity from the bar exam passage study. ``EM'' and ``MCMC'' estimates were computed using the \textbf{COMBO} R Package. The ``SAMBA'' results assume perfect specificity and were computed using functions from the \textbf{SAMBA} R Package. The ``PSens'' results were computed using an adapted function from the \textbf{SAMBA} R Package  \citep{SAMBA, beesley2020statistical}. The ``Naive'' results were obtained by running a simple logistic regression model for $Y^* \sim X$, assuming no misclassification in $Y^*$. Entries of $\boldsymbol{1}$ indicate that a probability was set at 1 for a given method.} \label{bar-probability-results-table}
\begin{threeparttable}
\begin{tabular}{crrrrr}
\hline
 & \multicolumn{5}{c}{$P(Y^* = 1 | Y = 1)$} \\
 \cline{2-6}
 Student Race & EM & MCMC & SAMBA & PSens & Naive   \\
\hline
 White & 1.000 & 0.994 & 1.000 & $\boldsymbol{1}$ & $\boldsymbol{1}$ \\
 Black & 0.997 & 0.993 & 0.981 & $\boldsymbol{1}$ & $\boldsymbol{1}$ \\
\hline
 & \multicolumn{5}{c}{$P(Y^* = 2 | Y = 2)$} \\
 \cline{2-6}
 Student Race & EM & MCMC & SAMBA & PSens & Naive   \\
\hline
 White & 0.475 & 0.373 & $\boldsymbol{1}$ & 0.485 & $\boldsymbol{1}$ \\
 Black & 0.562 & 0.492 & $\boldsymbol{1}$ & 0.577 & $\boldsymbol{1}$ \\
\hline 
\end{tabular}
\end{threeparttable}
\end{table}

The $\beta_2$ and $\beta_3$ estimates from all presented methods suggest that higher LSAT scores and law school GPAs, respectively, are associated with higher log odds of being bar admission. Notably, the parameter estimates from the naive model are attenuated compared to that of our proposed MCMC and EM procedures. This suggests that an analysis which ignores bar passage misclassification may underestimate the association between these student factors and the log odds of being admitted to the bar. 

In Table \ref{bar-probability-results-table}, we see that the EM, MCMC, and SAMBA methods all provide sensitivity estimates at or near 1.000. In contrast, $\boldsymbol{\gamma}$ estimates from EM, MCMC, and PSens approaches result in imperfect specificity estimates that vary with student race. In particular, we estimate that, among white students, $P(Y^* = 2 | Y = 2)$ is between 0.373 - 0.485 depending on the estimation method used. For Black students, we estimate $P(Y^* = 2 | Y = 2)$ values between 0.492 - 0.577 among the EM, MCMC, and PSens methods. It should be noted that the PSens modeling approach assumes perfect specificity in the observed outcome. In the presented scenario, both the EM and MCMC methods, which \textit{do not} assume perfect sensitivity, still result in near-perfect sensitivity estimates. This finding demonstrates that, even though our methods do not require perfect sensitivity or specificity assumptions, they can still handle cases where such an assumption may be appropriate.

In context, the sensitivity and specificity results suggest that the bar exam had near perfect sensitivity. That is, we estimate that individuals who should be admitted to the bar nearly \textit{always} pass the bar exam, regardless of race. However, among Black students who should not have been admitted to the bar, our EM results suggest a 56.2\% rate of bar exam failure. Among white students who should not have been admitted to the bar, our estimated failure rate from the EM algorithm is only 47.5\%. This result suggests that white students have a higher probability of spuriously passing the bar exam than Black students. These findings are in line with the original LSAC National Longitudinal Bar Passage Study, which found disparate bar passage rates white and Black examinees \citep{wightman1998lsac}.

\subsection[Evaluating the predictive accuracy of a risk assessment algorithm with COMBO]{Evaluating the predictive accuracy of a risk assessment algorithm with \textbf{COMBO}}\label{COMBO-2stage-example}

In this section, we detail a study pretrial decision-making using misclassification models. The goal of this study is to evaluate risk factors for the phenomenon of ``pretrial failure'', defined as instances where a defendant fails to appear for their trial date or reoffends before their trial date. In our study data, we have two potentially misclassified proxies for pretrial failure -- a risk assessment algorithm recommendation and a judicial decision. The risk assessment algorithm recommendation is obtained from the Virginia Pretrial Risk Assessment Instrument (VPRAI), which provides a recommendation to ``detain'' or ``release'' defendants awaiting trial based on several inputs \citep{lovins2016validation}. The judicial decision is the actual decision from a judge to release or detain an individual ahead of their trial. Judges generally consider the VPRAI recommendation when making their final decision, making this setting a case with sequential and dependent potentially misclassified outcome variables \citep{cadigan2011implementing, stevenson2018assessing, stevenson2022algorithmic, copp2022pretrial}. We hypothesize that both of these proxies for pretrial failure may be potentially misclassified based on the race of the defendant. In this study, we evaluate the predictive accuracy of our bias-corrected model for true pretrial failure compared to the VPRAI algorithm. This analysis is an extension of that presented in Webb, Riley, and Wells (2023)\cite{webb2023assessment}. 

In this study, we use pretrial data from admitted persons in Prince William County, Virginia between January 2016 and December 2019. The analysis only includes individuals for which a VPRAI recommendation was available. We also only included defendants whose race was recorded as white or Black in our dataset. The dataset contained 1,990 records after we removed all observations with missing values in the response variable and covariates of interest. Because of the sensitive nature of the data used for this study, we do not provide the original dataset in the \textbf{COMBO} R package. Instead, the \textbf{COMBO} R package contains a synthetic dataset that is similar to the original analysis data. In this paper, we show the code required to undertake the analysis and provide results from the real data (denoted \texttt{vprai\_data} in the example code). The synthetic dataset is used for the same example in Appendix \ref{vprai-synthetic-example}.

\subsubsection{Parameter estimation}
We wish to fit a two-stage misclassification model of the following form: 
\begin{equation}
\begin{aligned}
\label{eq:vprai-example-mechanisms}
\text{True outcome mechanism: } &\; \text{logit}\{ P(Y = 1 | \boldsymbol{X} ; \boldsymbol{\beta}) \} = \beta_{1} + \beta_2 \textsf{n\_FTA} + \beta_3 \textsf{unemployed} + \\
&\; \hspace{1.5em} \beta_4 \textsf{drug\_abuse} + \beta_5 \textsf{n\_violent\_arrest}. \\
\\
\text{ First-stage observation mechanisms: } &\; \text{logit}\{ P(\textsf{vprai} = 1 | Y = 1, Z^{(1)} ; \boldsymbol{\gamma}^{(1)}) \} \\ &\; \hspace{1.5em} = \gamma^{(1)}_{11} + \gamma^{(1)}_{21} \textsf{race},\\
&\; \text{logit}\{ P(\textsf{vprai} = 1 | Y = 2, Z^{(1)} ; \boldsymbol{\gamma}^{(1)}) \} \\ &\; \hspace{1.5em} = \gamma^{(1)}_{12} + \gamma^{(1)}_{22} \textsf{race}. \\
\\
\text{Second-stage observation mechanisms: } &\; \text{logit}\{ P(\textsf{judge} = 1 | \textsf{vprai} = 1, Y = 1, Z^{(2)} ; \boldsymbol{\gamma}^{(2)}) \}  \\ &\; \hspace{1.5em} = \gamma^{(2)}_{1111} + \gamma^{(2)}_{2111} \textsf{race},\\
&\; \text{logit}\{ P(\textsf{judge} = 1 | \textsf{vprai} = 2, Y = 1, Z^{(2)} ; \boldsymbol{\gamma}^{(2)}) \} \\ &\; \hspace{1.5em} = \gamma^{(2)}_{1121} + \gamma^{(2)}_{2121} \textsf{race},\\
&\; \text{logit}\{ P(\textsf{judge} = 1 | \textsf{vprai} = 1, Y = 2, Z^{(2)} ; \boldsymbol{\gamma}^{(2)}) \} \\ &\; \hspace{1.5em} = \gamma^{(2)}_{1112} + \gamma^{(2)}_{2112} \textsf{race},\\
&\; \text{logit}\{ P(\textsf{judge} = 1 |\textsf{vprai} = 2, Y = 2, Z^{(2)} ; \boldsymbol{\gamma}^{(2)}) \} \\ &\; \hspace{1.5em} = \gamma^{(2)}_{1122} + \gamma^{(2)}_{2122} \textsf{race}.
\end{aligned}
\end{equation}

In this example, $Y$ denotes the latent variable ``pretrial failure''. We model the probability of pretrial failure as a function of defendants' number of previous failures to appear (\texttt{n\_FTA}), employment status (\texttt{unemployed}; 0 = employed, 1 = unemployed), drug abuse history (\texttt{drug\_abuse}; 0 = no drug abuse history, 1 = history of drug abuse), and number of previous violent arrests (\texttt{n\_violent\_arrest}) in the \textit{true outcome mechanism}. In the \textit{first-stage observation mechanism}, the VPRAI recommendation (\texttt{vprai}; 0 = release, 1 = detain) is the first-stage potentially misclassified proxy for $Y$. We model the VPRAI recommendation, conditional on true pretrial failure status, as a function of the defendant's race (\texttt{race}; 0 = white, 1 = Black). In the \textit{second-stage observation mechanism}, the judge decision (\texttt{judge}; 0 = release, 1 = detain) is the second-stage potentially misclassified proxy for $Y$. The judge decision is modeled, conditional on the VPRAI recommendation and on $Y$, as a function of defendant race. 

We can fit the model in (\ref{eq:vprai-example-mechanisms}) using \textbf{COMBO} by first creating matrices for our predictors and observed outcomes of interest. 

\begin{lstlisting}
R> Ystar1 <- vprai_data$vprai
R> Ystar1_01 <- ifelse(Ystar1 == 2, 0, 1)
R> Ystar2 <- vprai_data$judge
R> Ystar2_01 <- ifelse(Ystar2 == 2, 0, 1)
R> x <- matrix(c(vprai_data$n_FTA, vprai_data$unemployed,
R>               vprai_data$drug_abuse, vprai_data$n_violent_arrest),
R>             ncol = 4, byrow = FALSE)
R> z1 <- matrix(c(vprai_data$race),
R>              ncol = 1, byrow = FALSE)
R> z2 <- matrix(c(vprai_data$race),
R>              ncol = 1, byrow = FALSE)
\end{lstlisting}

Next, we fit a naive regression model to obtain starting values for $\boldsymbol{\beta}$ terms in the EM algorithm. All $\boldsymbol{\gamma}$ terms are initiated at 0. 

\begin{lstlisting}
R> beta_start_glm <- glm(Ystar1_01 ~ x,
R>                       family = "binomial"(link = "logit"))
R> beta_start <- matrix(c(unname(coef(beta_start_glm))), ncol = 1)
R> gamma1_start <- matrix(rep(0, 4), ncol = 2)
R> gamma2_start <- array(rep(0, 8), dim = c(2,2,2))
\end{lstlisting}

The \texttt{COMBO\_EM\_2stage} function enables parameter estimation using our proposed EM algorithm as well as from a naive analysis model that does not account for misclassification in $Y^{*(1)}$ and $Y^{*(2)}$.

\begin{lstlisting}
R> vprai_EM <- COMBO_EM_2stage(Ystar1 = Ystar1,
R>                             Ystar2 = Ystar2,
R>                             x_matrix = x,
R>                             z1_matrix = z1, 
R>                             z2_matrix = z2,
R>                             beta_start = beta_start,
R>                             gamma1_start = gamma1_start,
R>                             gamma2_start = gamma2_start)
R> vprai_EM
R>          Parameter    Estimates           SE Convergence
R> 1           beta_1  -3.51150952   0.10536362        TRUE
R> 2           beta_2   1.22371749   0.21772007        TRUE
R> 3           beta_3   0.73224644   0.05654958        TRUE
R> 4           beta_4   1.96847059   0.12567035        TRUE
R> 5           beta_5   0.28049692   0.02233658        TRUE
R> 6        gamma1_11  -0.02922365   0.19869735        TRUE
R> 7        gamma1_21   1.84321037   0.37624760        TRUE
R> 8        gamma1_12 -20.27043135   0.60476265        TRUE
R> 9        gamma1_22  15.34058728   0.60296961        TRUE
R> 10     gamma2_1111   1.57553647   0.29757381        TRUE
R> 11     gamma2_2111   0.32735229   0.37410445        TRUE
R> 12     gamma2_1121   0.89206630   0.20263389        TRUE
R> 13     gamma2_2121  15.64485750 182.61786600        TRUE
R> 14     gamma2_1112  -6.49156713   1.85769165        TRUE
R> 15     gamma2_2112   9.82225108   1.85922310        TRUE
R> 16     gamma2_1122  -0.41755136   0.01826179        TRUE
R> 17     gamma2_2122   0.45947238   0.01826179        TRUE
R> 18    naive_beta_1   0.17026648 -21.98868200        TRUE
R> 19    naive_beta_2   0.13385788   7.66794400        TRUE
R> 20    naive_beta_3   0.15234988   4.43390128        TRUE
R> 21    naive_beta_4   0.16689478  10.44254600        TRUE
R> 22    naive_beta_5   0.03018705   8.48954536        TRUE
R> 23 naive_gamma2_11   0.31713516   4.96802796        TRUE
R> 24 naive_gamma2_21   0.38509961   0.91497180        TRUE
R> 25 naive_gamma2_12   0.06630521  -4.79237633        TRUE
R> 26 naive_gamma2_22   0.09708565   4.46178904        TRUE
\end{lstlisting}

In Appendix \ref{COMBO-MCMC-2stage-appendix}, we demonstrate parameter estimation for this model using a Markov Chain Monte Carlo (MCMC) approach via the \texttt{COMBO\_MCMC\_2stage} funciton in \textbf{COMBO}.

\subsubsection{Risk prediction}
Next, we use the $\boldsymbol{\beta}$ parameter estimates obtained from \texttt{COMBO\_EM\_2stage} to calculate estimated probabilities of pretrial failure for each observation in the dataset. We view these estimated probabilities as bias-corrected risk estimates of pretrial failure. We then compare the predictive accuracy of our misclassification model-based risk estimates with the VPRAI recommendations using two approaches: (1) a misclassification-adjusted receiver operating characteristic (ROC) analysis \citep{zawistowski2017corrected} and (2) an ROC analysis using only data on individuals who were, in fact, released ahead of their trial date. 

First, we will evaluate the predictive accuracy of our bias-corrected risk estimates using a misclassification-adjusted receiver operating characteristic (ROC) analysis \citep{zawistowski2017corrected}. For this analysis, we first use the \texttt{true\_classification\_prob} function in \textbf{COMBO} to compute bias-corrected estimates of pretrial failure probability for every observation. 

\begin{lstlisting}
R> beta_estimates <- matrix(vprai_EM$Estimates[1:5], ncol = 1)
R> p_pretrial_failure <- true_classification_prob(beta_estimates, x)
R> vprai_data$p_pretrial_failure <- p_pretrial_failure$Probability[which(
R>   p_pretrial_failure$Y == 1)]
\end{lstlisting}

Because we do not have observed pretrial failure outcomes for every subject in the dataset, we cannot perform a generic ROC analysis using the entire dataset. Instead, we adapt the misclassification-correction approach of Zawistowski et al. (2017)\cite{zawistowski2017corrected} to replace the observed outcomes with the conditional predictive probability of pretrial failure for each subject. This probability is calculated conditional on the observed and potentially misclassified proxies for pretrial failure, $Y^{*(1)}$ and $Y^{*(2)}$, as well as the observed covariates $X$ and $Z$. The form of this probability is identical to the E-Step weights for the EM algorithm developed in Webb and Wells (2023)\cite{webb2023assessment}, and can be computed using an internal \textbf{COMBO} function, \texttt{w\_j\_2stage}. To use \texttt{w\_j\_2stage}, we first restructure the matrix forms of \texttt{Ystar1}, \texttt{Ystar2}, \texttt{x}, \texttt{z1}, and \texttt{z2}. Using these matrices, we compute misclassification probabilities for the VPRAI and for judicial decisions. The restructured matrices and response probabilities are arguments in the \texttt{w\_j\_2stage} function. This function returns a matrix containing conditional predicted probability of pretrial failure for each observation in the first column and conditional predicted probability of \textit{no} pretrial failure in the second column.

\begin{lstlisting}
R> Ystar1_matrix <- matrix(c(Ystar1_01, 1 - Ystar1_01),
R>                         ncol = 2, byrow = FALSE)
R> Ystar2_matrix <- matrix(c(Ystar2_01, 1 - Ystar2_01),
R>                         ncol = 2, byrow = FALSE)
R> Z1_design <- matrix(c(rep(1, nrow(z1)), c(z1)),
R>                     ncol = 2, byrow = FALSE)
R> gamma1_estimates <- matrix(vprai_EM$Estimates[6:9],
R>                            ncol = 2, byrow = FALSE)
R> pistar1_matrix <- COMBO:::pistar_compute(gamma1_estimates, Z1_design,
R>                                          nrow(Z1_design), 2)
R> Z2_design <- matrix(c(rep(1, nrow(z2)), c(z2)),
R>                     ncol = 2, byrow = FALSE)
R> gamma2_estimates <- array(vprai_EM$Estimates[10:17], dim = c(2,2,2))
R> pistar2_array <- COMBO:::pitilde_compute(gamma2_estimates, Z2_design,
R>                                          nrow(Z2_design), 2)
R> X_design <- matrix(c(rep(1, nrow(x)), c(x)),
R>                    nrow = nrow(x), byrow = FALSE)
R> pi_matrix <- COMBO:::pi_compute(beta_estimates, X_design, 
R>                                 nrow(X_design), 2)
R> predictive_prob_pf <- COMBO:::w_j_2stage(Ystar1_matrix,
R>                                          Ystar2_matrix,
R>                                          pistar2_array,
R>                                          pistar1_matrix,
R>                                          pi_matrix, 
R>                                          nrow(Ystar1_matrix), 2)
R> vprai_data$predictive_prob_pf <- predictive_prob_pf[,1]
\end{lstlisting}

\begin{figure}[h!]
\begin{center}
\includegraphics[width = \textwidth]{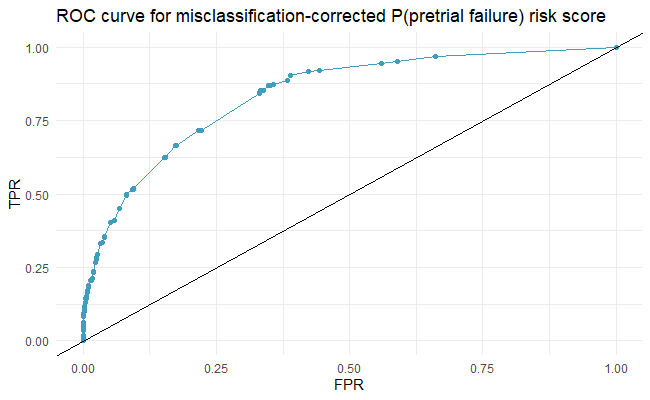}
\caption{Receiver-operating characteristic (ROC) curve for the misclassification-corrected predicted probability of pretrial failure from the VPRAI pretrial dataset. The ROC curve is obtained using the misclassification-adjusted ROC approach from \cite{zawistowski2017corrected}. FPR stands for false positive rate and TPR stands for true positive rate. }\label{misclassification-corrected-ROC}
\end{center}
\end{figure}

We use misclassification-adjusted true positive rate (TPR) and false positive rate (FPR) formulas from Zawistowski et al. (2017)\cite{zawistowski2017corrected} to construct an ROC curve (Figure \ref{misclassification-corrected-ROC}) and to obtain an AUC estimate. Using this method, the estimated AUC for our bias-corrected risk predictions is 0.838.

\begin{lstlisting}
R> cutoffs <- seq(0, 1, by = .01)
R> bias_corrected_TPR <- rep(NA, length(cutoffs))
R> bias_corrected_FPR <- rep(NA, length(cutoffs))
R> for(i in 1:length(cutoffs)){
R>   
R>   cutoff_i <- cutoffs[i]
R>   model_recommendation <- ifelse(vprai_data$p_pretrial_failure > cutoff_i,
R>                                  1, 0)
R>   
R>   bias_corrected_TPR[i] <- sum(predictive_prob_pf[,1] * 
R>                                  model_recommendation) /
R>                                sum(predictive_prob_pf[,1])
R>   
R>   bias_corrected_FPR[i] <- sum(predictive_prob_pf[,2] * 
R>                                  model_recommendation) /
R>                                sum(predictive_prob_pf[,2])  
R> }
R> 
R> bias_corrected_roc_data <- data.frame(TPR = bias_corrected_TPR,
R>                                       FPR = bias_corrected_FPR)
R> 
R> ggplot(data = model_roc_data) +
R>   geom_line(aes(x = FPR, y = TPR), color = "#409DBE") +
R>   geom_point(aes(x = FPR, y = TPR), color = "#409DBE") +
R>   theme_minimal() +
R>   geom_abline(slope = 1, intercept = 0) 
R> 
R> trapz(fliplr(bias_corrected_FPR), fliplr(bias_corrected_TPR))
R> [1] 0.8383721
\end{lstlisting}

In order to facilitate comparison between our bias-corrected model and the VPRAI, we conduct an ROC analysis using the subset of the data that has observed, ``ground truth'' pretrial failure outcomes. Such outcomes are, by design, only available for individuals who were released ahead of their trial date. Individuals who were detained ahead of their trial date do not have a pretrial failure outcome because they had no opportunity to fail to appear for their trial date or to reoffend before their trial date. By only evaluating our model predictions on individuals who were released by the court, we encounter the \textit{selective labels} problem \citep{coston2021characterizing, fogliato2020fairness, lakkaraju2017selective}. The selective labels problem generally refers to the bias resulting from analyses on a dataset that only contains outcomes for a subset of observations, due to a non-random selection mechanism. This issue is avoided when we use all data in the misclassification-adjusted ROC analysis, presented above \citep{zawistowski2017corrected}. However, the VPRAI cannot be evaluated using the misclassification-adjusted ROC analysis, because the conditional predictive probabilities used in place of an observed outcome are constructed using our misclassification model, not the VPRAI. 

To begin our predictive accuracy comparison for our model and the VPRAI, we first define the subset of observations from our dataset who were released ahead of their trial. 

\begin{lstlisting}
R> release_subset <- vprai_data %>%
R>   filter(judge == 2)
\end{lstlisting}

Next, we can compare bias-corrected model predictions between released individuals who did and did not experience a pretrial failure.

\begin{lstlisting}
R> release_subset %>%
R>   group_by(pretrial_failure) %>%
R>   summarise(avg_p_pretrial_failure = mean(p_pretrial_failure))
R> # A tibble: 2 x 2
R>   pretrial_failure avg_p_pretrial_failure
R>              <dbl>                  <dbl>
R> 1                0                  0.134
R> 2                1                  0.194
\end{lstlisting}

In this example, the average probability of pretrial failure among released individuals, computed from our misclassification model parameter estimates, is 0.194 among those who did experience a pretrial failure and 0.134 among those who did not experience a pretrial failure. 

\begin{figure}[h!]
\begin{center}
\includegraphics[width = \textwidth]{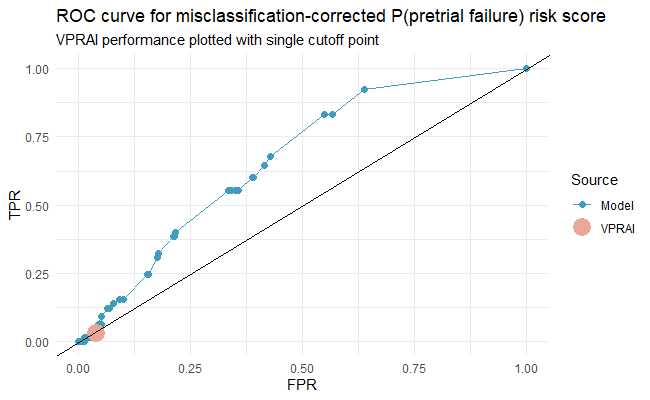}
\caption{Receiver-operating characteristic (ROC) curve for the misclassification-corrected predicted probability of pretrial failure from the VPRAI pretrial dataset, limited to released individuals who have gold standard pretrial failure outcomes. TPR and FPR of the VPRAI are included as a single point, since multiple cutoffs cannot be evaluated for the VPRAI algorithm (the VPRAI algorithm returns a ``release'' or ``detain'' recommendation, rather than a continuous risk prediction). FPR stands for false positive rate and TPR stands for true positive rate.}\label{vprai-roc}
\end{center}
\end{figure}

Next, we evaluate the accuracy of our model predictions and of VPRAI recommendations through an ROC analysis. For this analysis, bias-corrected predicted risk values are utilized (in place of naive risk predictions) to adjust for potential bias in ROC analyses from misclassified outcome variables. Because the VPRAI only provides a recommendation, rather than a risk probability, we essentially have one ``cutoff" at which to evaluate VPRAI accuracy. We include this evaluation as a single point in the constructed ROC curve (Figure \ref{vprai-roc}). The outcome variable used for this analysis is observed pretrial failure, denoted \texttt{pretrial\_failure} (0 = no pretrial failure, 1 = pretrial failure).

\begin{lstlisting}
R> cutoffs <- seq(0, 1, by = .01)
R> model_sensitivity <- rep(NA, length(cutoffs))
R> model_specificity <- rep(NA, length(cutoffs))
R> for(i in 1:length(cutoffs)){
R>   
R>   cutoff_i <- cutoffs[i]
R>   model_recommendation <- ifelse(release_subset$p_pretrial_failure > 
R>                                    cutoff_i,
R>                                  1, 0)
R>   
R>   model_sensitivity[i] <- (length(which(model_recommendation == 1 &
R>                                release_subset$pretrial_failure == 1))) /
R>     length(which(release_subset$pretrial_failure == 1))
R>   
R>   model_specificity[i] <- (length(which(model_recommendation == 0 &
R>                                release_subset$pretrial_failure == 0))) /
R>     length(which(release_subset$pretrial_failure == 0))  
R> }
R> 
R> VPRAI_sensitivity <- (length(which(release_subset$vprai == 1 &
R>                           release_subset$pretrial_failure == 1))) /
R>   length(which(release_subset$pretrial_failure == 1))
R>
R> VPRAI_specificity <- (length(which(release_subset$vprai == 2 &
R>                           release_subset$pretrial_failure == 0))) /
R>   length(which(release_subset$pretrial_failure == 0))
R> 
R> model_roc_data <- data.frame("FPR" = c(1 - model_specificity,
R>                                        1 - VPRAI_specificity),
R>                              "TPR" = c(model_sensitivity,
R>                                        VPRAI_sensitivity),
R>                              "Source" = c(rep("Model", 101),
R>                                           "VPRAI"))
R> 
R> ggplot(data = model_roc_data) +
R>   geom_line(aes(x = FPR, y = TPR, color = Source)) +
R>   geom_point(aes(x = FPR, y = TPR, color = Source, size = Source)) +
R>   theme_minimal() +
R>   geom_abline(slope = 1, intercept = 0) +
R>   scale_color_manual(values = c("#409DBE", "#ECA698"))
R> 
R> ## Model AUC
R> trapz(fliplr(1 - model_specificity), fliplr(model_sensitivity))
R> [1] 0.6682248
R> ## VPRAI AUC
R> release_subset$vprai_01 <- ifelse(release_subset$vprai == 1, 1, 0)
R> auc(release_subset$pretrial_failure, release_subset$vprai_01)
R> Setting levels: control = 0, case = 1
R> Setting direction: controls < cases
R> Area under the curve: 0.4962
\end{lstlisting}

Our misclassification-adjusted risk predictions achieve an AUC of 0.6682. In contrast, the AUC for the VPRAI recommendation algorithm is 0.4962 (to compute this, we consider the risk score equal to 1 if the VPRAI recommends detention and 0 otherwise). The AUC estimate from this analysis is notably lower than that obtained from the misclassification-adjusted procedure. However, both AUC estimates for the bias-corrected model-based approach are above 0.50 and both are above that of the VPRAI, suggesting that our model has improved predictive accuracy compared to chance and compared to the VPRAI.

\section[Methods]{Methods for misclassified binary mediator variables}\label{comma-methods}
In a mediation analysis, we are interested in the relationship between a predictor, $X$, and an outcome variable $Y$. $X$ may be directly associated with $Y$, and it may also be associated with $Y$ indirectly, through an intermediate or mediator variable $M$. In our setting, $M$ is a latent, binary variable taking values $j \in \{1,2\}$. Let $M^*$ denote an observed, but potentially misclassified version of $M$. $M^*$ is also a binary variable, taking on values $\ell \in \{1,2\}$. The matrix of predictors $\boldsymbol{Z}$ is related to the sensitivity and specificity of the mediator variable. We model these relationships as follows:
\begin{equation}
\begin{aligned}
\label{eq:comma-mechanism}
\text{True mediator mechanism: } &\; \logit\{ P(M = 1 | X, \boldsymbol{C} ; \boldsymbol{\beta}) \} = \beta_{0} + \beta_{X} X + \boldsymbol{\beta_{C} C}.\\
\text{Observed mediator mechanisms: } &\; \logit\{ P(M^* = 1 | M = 1, \boldsymbol{Z} ; \boldsymbol{\gamma}) \} = \gamma_{110} + \boldsymbol{\gamma_{11Z} Z}, \\
                                &\; \logit\{ P(M^* = 1 | M = 2, \boldsymbol{Z} ; \boldsymbol{\gamma}) \} = \gamma_{120} + \boldsymbol{\gamma_{12Z} Z}. \\
\text{Outcome mechanism: } &\;E(Y | X, \boldsymbol{C}, M; \boldsymbol{\theta}) = \theta_0 + \theta_X X + \boldsymbol{\theta_C C} + \theta_M M + \theta_{XM} XM,
\end{aligned}
\end{equation}
where $\boldsymbol{C}$ represents a matrix of continuous or categorical covariates. If $Y \sim Bernoulli$, we can replace the \textit{outcome mechanism} in (\ref{eq:comma-mechanism}) with a logit link, as follows:
\begin{equation}
\begin{aligned}
\label{eq:bernoulli-outcome-mechanism}
\logit \{P(Y = 1 | X, \boldsymbol{C}, M; \boldsymbol{\theta}) \} = \theta_0 + \theta_X X + \boldsymbol{\theta_C C} + \theta_m M + \theta_{XM} XM.
\end{aligned}
\end{equation}
Similarly, if $Y \sim Poisson$, we can replace replace the \textit{outcome mechanism} in (\ref{eq:comma-mechanism}) with a log link, as follows:
\begin{equation}
\begin{aligned}
\label{eq:poisson-outcome-mechanism}
\log \{E(Y | X, \boldsymbol{C}, M; \boldsymbol{\theta}) \} = \theta_0 + \theta_X X + \boldsymbol{\theta_C C} + \theta_m M + \theta_{XM} XM.
\end{aligned}
\end{equation}

Using (\ref{eq:comma-mechanism}), we can obtain expressions for the true mediator and observed mediator conditional on the true mediator response probabilities for all observations $i = 1, \dots, N$ in the sample:
\begin{align}
\label{eq:pi-m}
P(M_i = 1 | X_i = x, \boldsymbol{C_i} ; \boldsymbol{\beta}) = &\; \; \pi_{i1} = \frac{\text{exp}\{\beta_{0} + \beta_{X} x_i + \boldsymbol{\beta_{C} C_i} \} }{1 + \text{exp}\{\beta_{0} + \beta_{X} x_i + \boldsymbol{\beta_{C} C_i}\}}, \\
\label{pistar-m}
P(M^*_i = 1 | M_i = j, \boldsymbol{Z} ; \boldsymbol{\gamma}) = &\; \pi^*_{i1j} = \frac{\text{exp}\{\gamma_{1j0} + \boldsymbol{\gamma_{1jZ} Z_i}\}}{1 + \text{exp}\{\gamma_{1j0} + \boldsymbol{\gamma_{1jZ} Z_i}\}}.
\end{align}
Average sensitivity and specificity for the mediator are expressed as $\pi^*_{11} = \frac{1}{N}\sum_{i = 1}^{N} \pi^*_{i11}$ and $\pi^*_{22} = \frac{1}{N}\sum_{i = 1}^{N} \pi^*_{i22}$, respectively.

Our goal is to estimate parameters $(\boldsymbol{\beta}, \boldsymbol{\gamma}, \boldsymbol{\theta})$ in (\ref{comma-methods}). These parameter estimates can be used to compute estimates of controlled direct effects, indirect effects, and direct effects, as described in Webb and Wells (2024)\cite{webb2024effect}. In the following sections, we describe three estimation methods provided in the \textbf{COMMA} R package, including an ordinary least squares (OLS) correction, a predictive value weighting (PVW) procedure, and a seamless expectation-maximization (EM) algorithm.

\subsubsection{An ordinary least squares correction for misclassified binary mediator models}\label{comma-ols}
For the OLS correction approach, we first estimate $(\boldsymbol{\beta}, \boldsymbol{\gamma})$ using the EM algorithm described in \ref{combo1-em} and implemented in the \textbf{COMBO} R package. This method essentially replaces the latent and observed outcome variables from the original method with the latent and observed mediator variables $M$ and $M^*$ in this setting. We denote the estimates from the \textbf{COMBO} EM algorithm as $\hat{\boldsymbol{\beta}}^W$ and $\hat{\boldsymbol{\gamma}}^W$.

We extend the bias correction method from Nguimkeu, Rosenman, and Tennekoon (2021)\cite{nguimkeu2021regression} to estimate the $\boldsymbol{\theta}$ parameters, using estimated sensitivity and specificity rates. We estimate average mediator sensitivity and specificity across all subjects $i$ in the dataset using EM algorithm estimates as follows:
\begin{equation}
\begin{aligned}
\label{eq:avg_pistar}
\hat{\pi}^{*W}_{11} = \frac{1}{N}\sum_{i = 1}^N \hat{\pi}^{*W}_{i11} = \frac{1}{N}\sum_{i = 1}^N \frac{\text{exp}\{\hat{\gamma}^W_{110} + \boldsymbol{\hat{\gamma}^W_{11Z} Z_i}\}}{1 + \text{exp}\{\hat{\gamma}^W_{110} + \boldsymbol{\hat{\gamma}^W_{11Z} Z_i}\}},\\
\hat{\pi}^{*W}_{22} = \frac{1}{N}\sum_{i = 1}^N \hat{\pi}^{*W}_{i22} = \frac{1}{N}\sum_{i = 1}^N \frac{1}{1 + \text{exp}\{\hat{\gamma}^W_{120} + \boldsymbol{\hat{\gamma}^W_{12Z} Z_i}\}},
\end{aligned}
\end{equation}
where $\hat{\pi}^{*W}_{11}$ denotes estimated sensitivity and $\hat{\pi}^{*W}_{22}$ denotes estimated specificity.

Next, we compute bias-corrected $\boldsymbol{\theta}$ estimates as follows: 
\begin{equation}
\label{ols-correction-system}
\begin{bmatrix} \hat{\theta}_M  \\ \hat{\theta}_{D} \end{bmatrix} = \begin{bmatrix} (1 - \zeta) S_{M^*M^*} & S_{M^*D}  \\ (1 + \xi) S_{DM^*} & S_{DD}  \end{bmatrix}^{-1}  \begin{bmatrix} S_{YM^*}  \\ S_{YD} \end{bmatrix}.
\end{equation}
In (\ref{ols-correction-system}), $S_{AB} = \frac{1}{N} \sum_{i = 1}^N (A - \bar{A}) (B - \bar{B})^T$ denotes the sample covariance between variables $A$ and $B$, where $\bar{A}$ and $\bar{B}$ denote the sample means of $A$ and $B$, respectively. $\boldsymbol{D}$ is the matrix containing both the predictor of interest $X$ and the covariate matrix $\boldsymbol{C}$ from the \textit{outcome mechanism} in (\ref{eq:comma-mechanism}). The quantities $\zeta$ and $\xi$ are defined as,
\begin{equation}
\begin{aligned} \label{comma-squiggles}
     &\zeta = 1 - \frac{(\pi^*_1 - \hat{\pi}^{*W}_{12}) (1 - \hat{\pi}^{*W}_{21} - \pi^*_{1})}{(1 - \hat{\pi}^{*W}_{12} - \hat{\pi}^{*W}_{21}) (1 - \pi^*_1) \pi^*_1} \\
     &\xi = \frac{(\hat{\pi}^{*W}_{21} + \hat{\pi}^{*W}_{12})}{(1 - \hat{\pi}^{*W}_{12} - \hat{\pi}^{*W}_{21})},
\end{aligned}
\end{equation}
where $\hat{\pi}^{*W}_{21} = 1 - \hat{\pi}^{*W}_{11}$ and $\hat{\pi}^{*W}_{12} = 1 - \hat{\pi}^{*W}_{22}$. In (\ref{comma-squiggles}), $\pi^*_1 = P(M^*_i = 1)$ is the empirical response probability of the observed mediator. 

Finally, the intercept in the \textit{outcome mechanism} in (\ref{eq:comma-mechanism}) is computed as: 
\begin{equation}
    \hat{\theta}_0 = \bar{Y} - \hat{\theta}_M \frac{\bar{M}^* - \hat{\pi}^{*W}_{12}}{(1 - \hat{\pi}^{*W}_{12} - \hat{\pi}^{*W}_{21})} - \bar{D}^T\hat{\theta}_D.
\end{equation}

It should be noted that the OLS correction method currently cannot support an interaction term in the \textit{outcome mechanism} in (\ref{eq:comma-mechanism}). Moreover, the OLS correction method is only appropriate for approximately normally distributed outcome variables. 

\subsubsection{A predictive value weighting procedure for misclassified binary mediator models}\label{comma-pvw}
Much like the OLS correction method, we begin the predictive value weighting (PVW) procedure by estimating $(\boldsymbol{\beta}, \boldsymbol{\gamma})$ via the \textbf{COMBO} EM algorithm approach. This method yields parameter estimates $\hat{\boldsymbol{\beta}}^W$ and $\hat{\boldsymbol{\gamma}}^W$. We extend the PVW approach from Lyles and Lin (2010)\cite{lyles2010sensitivity} by estimating correct classification rates from the data, rather than using known values. We estimate mediator sensitivity and specificity, $\hat{\pi}^*_{i11}$ and $\hat{\pi}^*_{i22}$, respectively, for all observations $i = 1, \dots, N$, as follows: 
\begin{equation}
\begin{aligned}
\label{eq:pistar-pvw-est}
\hat{\pi}^{*W}_{i11} = \frac{\text{exp}\{\hat{\gamma}^W_{110} + \boldsymbol{\hat{\gamma}^W_{11Z} Z_i}\}}{1 + \text{exp}\{\hat{\gamma}^W_{110} + \boldsymbol{\hat{\gamma}^W_{11Z} Z_i}\}},\\
\hat{\pi}^{*W}_{i22} = \frac{1}{1 + \text{exp}\{\hat{\gamma}^W_{120} + \boldsymbol{\hat{\gamma}^W_{12Z} Z_i}\}}.
\end{aligned}
\end{equation}

Next, we fit a logistic regression model for the observed mediator variable $M^*$ as a function of other observed variables $Y$, $X$, and $\boldsymbol{C}$. The \textbf{COMMA} R package automatically includes interaction terms between all observed variables in the model. We use the parameter estimates from this logistic regression model to compute estimated observed mediator response probabilities $P(M^*_i = 1 | Y_i, X_i, \boldsymbol{C_i})$. Using these response probabilities, as well as the subject-level sensitivity and specificity estimates from (\ref{eq:pistar-pvw-est}), we compute subject-level positive predictive value (PPV) and negative predictive value (NPV) estimates:
\begin{equation}
\begin{aligned}
\label{comma-ppv-and-npv}
    &PPV_i = \frac{\Bigl[ \frac{(\hat{\pi}^{*W}_{i22} - 1) \times (P(M^*_i = 1 | Y_i, X_i, \boldsymbol{C_i}) - 1)}{\hat{\pi}^{*W}_{i22} \times P(M^*_i = 1 | Y_i, X_i, \boldsymbol{C_i})} - 1 \Bigr]}{\Bigl[ \Bigl( \frac{(\hat{\pi}^{*W}_{i11} - 1) \times P(M^*_i = 1 | Y_i, X_i, \boldsymbol{C_i})}{\hat{\pi}^{*W}_{i11} \times (P(M^*_i = 1 | Y_i, X_i, \boldsymbol{C_i}) - 1)} \Bigr) \Bigl( \frac{(\hat{\pi}^{*W}_{i22} - 1) \times (P(M^*_i = 1 | Y_i, X_i, \boldsymbol{C_i}) - 1)}{\hat{\pi}^{*W}_{i22} \times P(M^*_i = 1 | Y_i, X_i, \boldsymbol{C_i})} \Bigr) - 1 \Bigr]} \\
    &NPV_i = \frac{\Bigl[ \frac{(\hat{\pi}^{*W}_{i11} - 1) \times P(M^*_i = 1 | Y_i, X_i, \boldsymbol{C_i})}{\hat{\pi}^{*W}_{i11} \times (P(M^*_i = 1 | Y_i, X_i, \boldsymbol{C_i}) - 1)} - 1 \Bigr]}{\Bigl[ \Bigl( \frac{(\hat{\pi}^{*W}_{i11} - 1) \times P(M^*_i = 1 | Y_i, X_i, \boldsymbol{C_i})}{\hat{\pi}^{*W}_{i11} \times (P(M^*_i = 1 | Y_i, X_i, \boldsymbol{C_i}) - 1)} \Bigr) \Bigl( \frac{(\hat{\pi}^{*W}_{i22} - 1) \times (P(M^*_i = 1 | Y_i, X_i, \boldsymbol{C_i}) - 1)}{\hat{\pi}^{*W}_{i22} \times P(M^*_i = 1 | Y_i, X_i, \boldsymbol{C_i})} \Bigr) - 1 \Bigr]}.
\end{aligned}
\end{equation}
Next, we duplicate the original dataset. In both the original and duplicated datasets, we add a column for the true mediator variable $M$. In the original dataset, we set $M = 1$. In the duplicated dataset, $M = 0$. Stacking the datasets together, we create a ``weight'' column $W$ according to the following pattern:
\begin{equation}
\begin{aligned}
    &M_i = 1 \cap M^*_i = 1 \implies W_i = PPV_i \\
    &M_i = 0 \cap M^*_i = 1 \implies W_i = 1 - PPV_i \\
    &M_i = 1 \cap M^*_i = 0 \implies W_i = 1 - NPV_i \\
    &M_i = 0 \cap M^*_i = 0 \implies W_i = NPV_i.
\end{aligned}
\end{equation}

To estimate $\boldsymbol{\theta}$, we use the new, combined dataset to fit a weighted regression model for $Y | M, X, \boldsymbol{C}$, with weights equal to $W$. The \textbf{COMMA} R package can support $Y$ distributed as Normal, Bernoulli, and Poisson, and the appropriate generalized linear regression is fit to estimate $\boldsymbol{\theta}$, depending on the user-input distribution in the \textbf{COMMA} PVW functions. \textbf{COMMA} PVW functions can also support \textit{outcome mechanism} formulations both with and without an interaction term, $\theta_{XM}$.

\subsubsection{An EM algorithm for misclassified binary mediator models}\label{comma-em}
In this section, we provide a seamless EM algorithm approach for estimating $(\boldsymbol{\beta}, \boldsymbol{\gamma}, \boldsymbol{\theta}$) in (\ref{eq:comma-mechanism}). First, we write the complete data log-likelihood:
\begin{equation}
    \begin{aligned}
    \label{eq:complete-log-like-em}
    \ell_{complete}(\boldsymbol{\beta},& \boldsymbol{\gamma}, \boldsymbol{\theta}; X, \boldsymbol{C}, \boldsymbol{Z}, Y) \\
    &= \sum_{i = 1}^N \bigg[ \ell_{Y | X, M, \boldsymbol{C}}(\boldsymbol{\theta}; X_i, M_i, \boldsymbol{C}_i, Y_i) +
    \ell_{M | X, \boldsymbol{C}}(\boldsymbol{\beta}; X_i, M_i, \boldsymbol{C}_i) +
    \ell_{M^* | M, \boldsymbol{Z}}(\boldsymbol{\gamma};M_i, \boldsymbol{Z}_i, M^*_i) \bigg] & \\
    &= \sum_{i = 1}^N \bigg[ \ell_{Y | X, M, \boldsymbol{C}}(\boldsymbol{\theta}; X_i, M_i, \boldsymbol{C}_i, Y_i) + \sum_{j = 1}^2 m_{ij} \text{log} \{ \pi_{ij} \} + \sum_{j = 1}^2 \sum_{\ell = 1}^2  m_{ij} m^*_{i\ell} \text{log} \{ \pi^*_{i \ell j} \}\bigg],
    \end{aligned}
\raisetag{12pt}\end{equation}
where $m_{ij} = \mathbbm{I}(M_i = j)$ and $m^*_{i\ell} = \mathbbm{I}(M^*_i = \ell)$. A generic likelihood $\ell_{Y | X, M, \boldsymbol{C}}(\boldsymbol{\theta}; X_i, M_i, \boldsymbol{C}_i, Y_i)$ is included for the outcome (Normal, Bernoulli, and Poisson) specifications. This choice introduces a slight abuse of notation. In (\ref{eq:complete-log-like-em}), $\boldsymbol{\theta}$ now represents a vector of all \textit{outcome mechanism} parameters, including the variance $\sigma$, for example, if we model $Y$ as a Normal distribution.

For the E-step, we compute the expectation of $m_{ij}$ as
\begin{equation}
\begin{aligned}
\label{eq:e-step-em}
w_{ij} = P(M_i = j | M_i^*, X_i, \boldsymbol{C}_i, \boldsymbol{Z}_i, Y_i) = \sum_{\ell = 1}^2 \frac{m^*_{i\ell} \pi^*_{i\ell j} \pi_{ij} E[Y_i | X_i, M_i = j, \boldsymbol{C}_i, \boldsymbol{\theta}^{(t)}]}{\sum_{k = 1}^2 \pi^*_{i \ell k} \pi_{ik}E[Y_i | X_i, M_i = k, \boldsymbol{C}_i, \boldsymbol{\theta}^{(t)}]},
\end{aligned}
\end{equation}
where $t$ denotes the iteration of the EM algorithm.

Because (\ref{eq:complete-log-like-em}) is linear in the latent variable $m_{ij}$, we can replace $m_{ij}$ in (\ref{eq:complete-log-like-em}) with its expectation to construct the $Q$ function for the M-step:
\begin{equation}
    \begin{aligned}
    \label{eq:q-em}
    Q
    = \sum_{i = 1}^N \big[ \sum_{j = 1}^2 \ell_{Y | X, M, \boldsymbol{C}}(\boldsymbol{\theta}; X_i, M_i = w_{ij}, \boldsymbol{C}_i, Y_i) + \sum_{j = 1}^2 w_{ij} \text{log} \{ \pi_{ij} \} + \sum_{j = 1}^2 \sum_{\ell = 1}^2  w_{ij} m^*_{i\ell} \text{log} \{ \pi^*_{i \ell j} \} \big].
    \end{aligned}
\end{equation}
The $Q$ function can again be expressed as separate functions, containing parameters $\boldsymbol{\beta}$, $\boldsymbol{\gamma_1}$, $\boldsymbol{\gamma_2}$, and $\boldsymbol{\theta}$, respectively:
\begin{equation}
\begin{aligned}
\label{eq:comma-q-param}
 &Q_{\boldsymbol{\beta}} = \sum_{i = 1}^N \big[ \sum_{j = 1}^2 w_{ij} \text{log} \{ \pi_{ij} \} \big], \\
 &Q_{\boldsymbol{\gamma}_{1}} = \sum_{i = 1}^N \big[ \sum_{\ell = 1}^2  w_{i1}  m^*_{i\ell} \text{log} \{ \pi^*_{i \ell 1} \} \big], 
Q_{\boldsymbol{\gamma}_{2}} = \sum_{i = 1}^N \big[ \sum_{\ell = 1}^2  w_{12}  m^*_{i\ell} \text{log} \{ \pi^*_{i \ell 2} \} \big], \\
    &Q_{\boldsymbol{\theta}}
    = \sum_{i = 1}^N \big[ \sum_{j = 1}^2 \ell_{Y | X, M, \boldsymbol{C}}(\boldsymbol{\theta}; X_i, M_i = w_{ij}, \boldsymbol{C}_i, Y_i) \big].
\end{aligned}
\end{equation}
In the \textbf{COMMA} R package, we fit $Q_{\boldsymbol{\beta}}$ in (\ref{eq:comma-q-param}) as a standard logistic regression model where the outcome variable is $w_{ij}$. $Q_{\boldsymbol{\gamma}_{1}}$ and $Q_{\boldsymbol{\gamma}_{2}}$ in (\ref{eq:comma-q-param}) are fit as weighted logistic regression models, with weights equal to $w_{i1}$ or $w_{i2}$, respectively, and with outcome variable $m^*_{i\ell}$.

When the user specifies that $Y \sim Normal$, the \textbf{COMMA} R package estimates $Q_{\boldsymbol{\theta}}$ using traditional maximum likelihood methods. When the user specifies that $Y$ is distributed as a Bernoulli or Poisson, we estimate $Q_{\boldsymbol{\theta}}$ through a ``duplicated data'' approach. The dataset is duplicated and stacked. Then, a column is added for the true mediator $M$. $M = 0$ in the original dataset and $M = 1$ in the duplicated dataset. We also add a weight column $W^{EM}$ such that $W^{EM} = w_{i1}$ when $M_i = 1$ and $W^{EM} = w_{i2}$ when $M_i = 0$. To estimate $Q_{\boldsymbol{\theta}}$, we fit a weighted generalized linear model to the duplicated dataset, using the logit or log link for $Y$ distributed as a Bernoulli or Poisson, respectively. 

In order to obtain the final parameters estimates for $\boldsymbol{\beta}$, $\boldsymbol{\gamma}$, and $\boldsymbol{\theta}$ using the EM algorithm, the \textbf{COMMA} R package applies a label switching correction. This correction is described in more detail in Webb and Wells (2024)\cite{webb2024effect}.

\textbf{COMMA} EM algorithm functions can support \textit{outcome mechanism} formulations both with and without an interaction term, $\theta_{XM}$.

\section[The COMMA package]{The \textbf{COMMA} package}\label{comma-example}

\subsection[Installation and basic usage]{Installation and basic usage}
The \textbf{COMMA} package can be downloaded from CRAN. More information on the package can also be found at \href{https://cran.r-project.org/web/packages/COMMA/index.html}{https://cran.r-project.org/web/packages/COMMA/index.html}. Installing and loading \textbf{COMMA} in \texttt{R} requires the following code:
\begin{lstlisting}
R> install.packages("COMMA")
R> library(COMMA)
\end{lstlisting}

\subsection[Investigating the mediating role of gestational hypertension in the relationship between ethnicity and birthweight using COMMA]{Investigating the mediating role of gestational hypertension in the relationship between ethnicity and birthweight using \textbf{COMMA}}
We demonstrate usage of the \textbf{COMMA} package through a study of the birthweight outcomes among Hispanic and non-Hispanic White mothers. We investigate the potential mediating role of gestational hypertension on the relationship between ethnicity and birthweight, replicating a similar analysis performed by Li and VanderWeele (2020)\cite{li_direct_2020}. For this case study, we use data from the National Vital Statistics System of the National Center for Health Statistics (NCHS) from the year 2022 \citep{NCHS}. We limit the dataset to include nulliparous mothers with singleton births who are not missing data on our exposure, outcome, mediator, or covariates of interest. To manage computation time, we perform our analysis on a random subset of the data of size $n = 20,000$. The sample data is available is provided in \textbf{COMMA} and can be loaded using the following code: 
\begin{lstlisting}
R> data("NCHS2022_sample")
\end{lstlisting}

\subsubsection{Parameter estimation with a continuous outcome}
\begin{table}[htbt]
\centering
\caption{Variables of interest in the mediation analysis from section \ref{comma-example}. In this table, we include the following abbreviations: GH = gestational hypertension, BMI = body mass index, GD = gestational diabetes.} \label{COMMA-vars-table}
\begin{threeparttable}
\begin{tabular}{crrrrr}
 Notation & Definition & Variable & Description \\
\hline
 $X$ & Exposure & \textsf{ethnicity} & \makecell[r]{Indicator variable; \\ 0 = Non-Hispanic white, \\ 1 = Hispanic} \\[.75cm]
 $M$ & Latent Mediator & - & \makecell[r]{Indicator variable; \\ 0 = No GH, 1 = GH} \\[.5cm]
 $M^*$ & Observed Mediator & \textsf{g\_hyp} & \makecell[r]{Indicator variable;\\ 0 = No GH, 1 = GH}\\[.5cm]
 $Y$ & Outcome & \textsf{birth\_wt} & \makecell[r]{Continuous variable;\\ Birthweight (kilograms)}\\[.5cm]
 $C_1$ & Covariate & \textsf{age\_cs} & \makecell[r]{Continuous variable;\\ Mother's age (centered, scaled)}\\[.5cm]
 $C_2$ & Covariate & \textsf{pre\_hyp} & \makecell[r]{Indicator variable; \\ 0 = No pre-pregnancy hypertension, \\ 1 = Pre-pregnancy hypertension}\\[.75cm]
 $C_3$ & Covariate & \textsf{bmi\_cs} & \makecell[r]{Continuous variable; \\ Mother's BMI (centered, scaled)}\\[.5cm]
 $C_4$ & Covariate & \textsf{cig\_use} & \makecell[r]{Indicator variable; \\ 0 = No smoking history, \\ 1 = Smoking history}\\[.75cm]
 $C_5$ & Covariate & \textsf{prenatal\_m} & \makecell[r]{Continuous variable; \\ Month of pregnancy when prenatal care began}\\[.5cm]
 $C_6$ & Covariate & \textsf{pre\_diab} & \makecell[r]{Indicator variable; \\ 0 = No pre-pregnancy diabetes, \\ 1 = Pre-pregnancy diabetes}\\[.75cm]
 $C_7$ & Covariate & \textsf{g\_diab} & \makecell[r]{Indicator variable; \\ 0 = No GD, 1 = GD}\\[.5cm]
 $Z_1$ & \makecell[r]{Misclass. Predictor} & \textsf{ethnicity} & \makecell[r]{Indicator variable; \\ 0 = Non-Hispanic white, \\ 1 = Hispanic} \\[.75cm]
 $Z_2$ & \makecell[r]{Misclass. Predictor} & \textsf{self\_pay} & \makecell[r]{Indicator variable; \\ 0 = Private ins. or Medicaid, \\ 1 = Self-Pay} \\[.75cm]
 $Z_3$ & \makecell[r]{Misclass. Predictor} & \textsf{prenatal\_m} & \makecell[r]{Continuous variable; \\ Month of pregnancy when prenatal care began}\\[.5cm]
 
\hline 
\end{tabular}
\end{threeparttable}
\end{table}

Table \ref{COMMA-vars-table} lists the variables of interest in this analysis. Using these variables, our goal is to fit the following model using the provided sample data. The \textit{true mediator mechanism} is defined as:
\begin{equation}
\begin{aligned}
\label{eq:COMMA-ex-true-mech}
\logit\{ P(M = 1 | X, \boldsymbol{C} ; \boldsymbol{\beta}) \} =& \beta_{0} + \beta_{1} \textsf{ethnicity} + \beta_{2} \textsf{age\_cs} + \beta_{3} \textsf{pre\_hyp} + \beta_{4} \textsf{bmi\_cs} + \beta_{5} \textsf{cig\_use} +\\
&\beta_{6} \textsf{prenatal\_m} + \beta_{7} \textsf{pre\_diab} + \beta_{8} \textsf{g\_diab}.
\end{aligned}
\end{equation}

For the \textit{observed mediator mechanisms}, we fit the following models:
\begin{equation}
\begin{aligned}
\label{eq:COMMA-ex-obs-mech}
\logit\{ P(\textsf{g\_hyp} = 1 | M = 1, \boldsymbol{Z} ; \boldsymbol{\gamma}) \} =& \gamma_{11} + \gamma_{21} \textsf{ethnicity} + \gamma_{31} \textsf{self\-pay} + \gamma_{41} \textsf{prenatal\_m}, \\
\logit\{ P(\textsf{g\_hyp} = 1 | M = 2, \boldsymbol{Z} ; \boldsymbol{\gamma}) \} =& \gamma_{12} + \gamma_{22} \textsf{ethnicity} + \gamma_{32} \textsf{self\-pay} + \gamma_{42} \textsf{prenatal\_m}.
\end{aligned}
\end{equation}

Finally, we consider the following \textit{outcome mechanism}:
\begin{equation}
\begin{aligned}
\label{eq:COMMA-ex-out-mech}
E(\textsf{birth\_wt} | X, \boldsymbol{C}, M; \boldsymbol{\theta}) =& \theta_{0} + \theta_{x1} \textsf{ethnicity} + \theta_{c1} \textsf{age\_cs} + \theta_{c2} \textsf{pre\_hyp} + \theta_{c3} \textsf{bmi\_cs} + \\ &\theta_{c4} \textsf{cig\_use} +
\theta_{c5} \textsf{prenatal\_m} + \theta_{c6} \textsf{pre\_diab} + \theta_{c7} \textsf{g\_diab} +\\ & \theta_m M + \theta_{xm} (\textsf{ethnicity} \times M).
\end{aligned}
\end{equation}

Before fitting these models using \textbf{COMMA}, we restructure our observed data by converting the coding of the \texttt{g\_hyp} variable and creating matrices containing $X$, $C$, and $Z$. 
\begin{lstlisting}
R> bw_mstar <- matrix(ifelse(NCHS2022_sample$g_hyp == 1, 1, 2),
R>                    ncol = 1)
R> bw_mstar_01 <- NCHS2022_sample$g_hyp
R> bw_z <- matrix(c(NCHS2022_sample$ethnicity,
R>                  NCHS2022_sample$self_pay,
R>                  NCHS2022_sample$prenatal_m),
R>                nrow = nrow(NCHS2022_sample), byrow = FALSE)
R> bw_x <- matrix(NCHS2022_sample$ethnicity, ncol = 1)
R> bw_c <- matrix(c(NCHS2022_sample$age_cs,
R>                  NCHS2022_sample$pre_hyp,
R>                  NCHS2022_sample$bmi_cs,
R>                  NCHS2022_sample$cig_use,
R>                  NCHS2022_sample$prenatal_m,
R>                  NCHS2022_sample$pre_diab,
R>                  NCHS2022_sample$g_diab),
R>                nrow = nrow(NCHS2022_sample), byrow = FALSE)
R> bw_y <- NCHS2022_sample$birth_wt
\end{lstlisting}
The proposed approaches require starting values for the parameters. We start all $\boldsymbol{\gamma}$ parameters at 0 and run regression models that assume no misclassification in $M^*$ to obtain starting values for $\boldsymbol{\beta}$ and $\boldsymbol{\theta}$.
\begin{lstlisting}
R> bw_beta_glm <- glm(bw_mstar_01 ~ bw_x + bw_c, 
R>                    family = "binomial"(link = "logit"))
R> bw_beta_start <- matrix(unname(coef(bw_beta_glm)), ncol = 1)
R> bw_gamma_start <- matrix(rep(0, 8), ncol = 2)
R> bw_theta_lm <- lm(bw_y ~ bw_x + bw_mstar_01 + bw_c + bw_x:bw_mstar_01)
R> bw_theta_start <- matrix(unname(coef(bw_theta_lm)))
\end{lstlisting}

First, we estimate model parameters using our proposed EM algorithm in the \texttt{COMMA\_EM} function.
\begin{lstlisting}
R> bw_EM <- COMMA_EM(Mstar = bw_mstar[,1],
R>                   outcome = bw_y,
R>                   outcome_distribution = "Normal",
R>                   interaction_indicator = TRUE,
R>                   x_matrix = bw_x,
R>                   z_matrix = bw_z,
R>                   c_matrix = bw_c,
R>                   beta_start = bw_beta_start,
R>                   gamma_start = bw_gamma_start,
R>                   theta_start = bw_theta_start,
R>                   sigma_start = sigma(bw_theta_lm),
R>                   tolerance = 1e-7,
R>                   max_em_iterations = 1500,
R>                   em_method = "squarem")
R> bw_EM
R>    Parameter    Estimates Convergence
R> 1     beta_0 -0.486960122        TRUE
R> 2     beta_1 -0.070020347        TRUE
R> 3     beta_2 -0.005474966        TRUE
R> 4     beta_3 -0.091760193        TRUE
R> 5     beta_4  0.053590115        TRUE
R> 6     beta_5  0.016364043        TRUE
R> 7     beta_6 -0.002744413        TRUE
R> 8     beta_7  0.298009581        TRUE
R> 9     beta_8  0.195860448        TRUE
R> 10   gamma11 -1.327033845        TRUE
R> 11   gamma21 -0.345561042        TRUE
R> 12   gamma31 -0.620198310        TRUE
R> 13   gamma41 -0.026292929        TRUE
R> 14   gamma12 -2.029829368        TRUE
R> 15   gamma22 -0.213619193        TRUE
R> 16   gamma32 -1.316177887        TRUE
R> 17   gamma42 -0.074815482        TRUE
R> 18   theta_0  3.654293637        TRUE
R> 19   theta_x -0.089875252        TRUE
R> 20   theta_m -0.813770530        TRUE
R> 21  theta_c1  0.005667262        TRUE
R> 22  theta_c2 -0.317411172        TRUE
R> 23  theta_c3  0.046439529        TRUE
R> 24  theta_c4 -0.226395512        TRUE
R> 25  theta_c5 -0.011583123        TRUE
R> 26  theta_c6 -0.068317162        TRUE
R> 27  theta_c7  0.016003577        TRUE
R> 28  theta_xm -0.014069626        TRUE
R> 29     sigma  0.126019966        TRUE
\end{lstlisting}

We obtain standard error (SE) estimates via the bootstrap using the \texttt{COMMA\_EM\_bootstrap\_SE} function. Note that only 100 bootstrap samples were generated in order to maintain reasonable computation time. As such, these results should be interpreted as a demonstration of \texttt{COMMA} only. Many more bootstrap samples (i.e. 2,000 or more) should be computed in order to draw conclusions from this study. 
\begin{lstlisting}
R> bw_EM_SE <- COMMA_EM_bootstrap_SE(parameter_estimates = 
R>                                      bw_EM$Estimates[-29],
R>                                   sigma_estimate = 
R>                                      bw_EM$Estimates[29],
R>                                   n_bootstrap = 100,
R>                                   n_parallel = 8,
R>                                   outcome_distribution = "Normal",
R>                                   interaction_indicator = TRUE,
R>                                   x_matrix = bw_x,
R>                                   z_matrix = bw_z,
R>                                   c_matrix = bw_c,
R>                                   tolerance = 1e-7,
R>                                   max_em_iterations = 1500,
R>                                   em_method = "squarem")
R> print(bw_EM_SE$bootstrap_SE[,c("Parameter", "SE")], n = 29)
R> # A tibble: 29 x 2
R>    Parameter      SE
R>    <chr>       <dbl>
R>  1 beta_0    0.0617 
R>  2 beta_1    0.0852 
R>  3 beta_2    0.0311 
R>  4 beta_3    0.221  
R>  5 beta_4    0.0289 
R>  6 beta_5    0.187  
R>  7 beta_6    0.0179 
R>  8 beta_7    0.304  
R>  9 beta_8    0.123  
R> 10 gamma11   0.0648 
R> 11 gamma12   0.0713 
R> 12 gamma21   0.0755 
R> 13 gamma22   0.0859 
R> 14 gamma31   0.224  
R> 15 gamma32   0.329  
R> 16 gamma41   0.0192 
R> 17 gamma42   0.0226 
R> 18 sigma     0.00611
R> 19 theta_0   0.0310 
R> 20 theta_c1  0.0144 
R> 21 theta_c2  0.110  
R> 22 theta_c3  0.0149 
R> 23 theta_c4  0.0952 
R> 24 theta_c5  0.00874
R> 25 theta_c6  0.157  
R> 26 theta_c7  0.0615 
R> 27 theta_m   0.0101 
R> 28 theta_x   0.0393 
R> 29 theta_xm  0.0241
\end{lstlisting}

Parameter estimates and SEs can be obtained using the PVW and OLS correction methods with the \texttt{COMMA\_PVW}, \texttt{COMMA\_PVW\_bootstrap\_SE}, \texttt{COMMA\_OLS}, and \texttt{COMMA\_OLS\_bootstrap\_SE} functions, respectively. Details on these functions are available in Appendix \ref{COMMA-PVS-OLS-appendix}. Figure \ref{COMMA_parameter_plot} displays the parameter estimates and SEs for all proposed methods as well as the naive approach that was used to obtain starting values. 

\begin{figure}[h!]
\begin{center}
\includegraphics[width = 400pt]{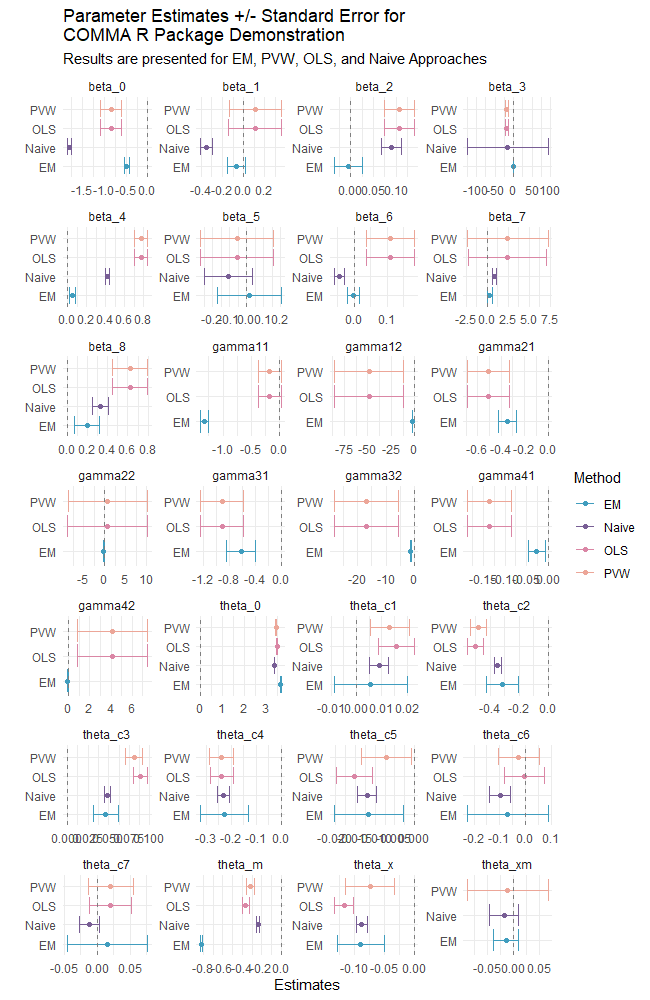}
\caption{(Caption next page.) }\label{COMMA_parameter_plot}
\end{center}
\end{figure}

\addtocounter{figure}{-1}

\begin{figure}
    \caption{(Previous page.) Parameter estimates +/- standard error from the \textbf{COMMA} R package demonstration. $\boldsymbol{\beta}$, $\boldsymbol{\gamma}$, and $\boldsymbol{\theta}$ parameters are introduced in (\ref{eq:COMMA-ex-true-mech}), (\ref{eq:COMMA-ex-obs-mech}), and (\ref{eq:COMMA-ex-out-mech}), respectively. Estimates are presented for the proposed expectation-maximization (EM) algorithm, predictive value weighting (PVW), and ordinary least squares (OLS) correction approaches, as well as a naive mediation analysis that assumes no misclassification in the mediator variable.}
\end{figure}

Effect estimates can be computed using the resulting $(\boldsymbol{\beta}, \boldsymbol{\gamma})$ parameters from each method. In Table \ref{comma-bw-effect-table}, we compute natural direct effects, natural indirect effects, and controlled direct effects following the formulas in Valeri and VanderWeele (2013)\cite{valeri2013mediation}. 

The $\boldsymbol{\beta}$ and $\boldsymbol{\theta}$ parameter estimates from the proposed EM algorithm are generally attenuated when compared to the naive mediation analysis that does not account for misclassification (Figure \ref{COMMA_parameter_plot}). A notable exception is the $\theta_m$ term, which is larger in magnitude when estimated via the EM algorithm than with any other method. This suggests that, after accounting for misclassification through our proposed EM algorithm, gestational hypertension has a stronger, negative impact on birthweight than what would have been estimated via a traditional mediation analysis. The PVW and OLS correction approaches tend to yield more variable parameter estimates, though with the small number of bootstrap samples drawn, conclusions are difficult to draw. It should be noted that the OLS correction approach does not support interaction term estimation in the \textit{outcome mechanism}, and so  $\theta_{xm}$ is not provided using this method.  In Table \ref{comma-bw-effect-table}, misclassification of gestational hypertension appears to have had little impact on effect estimates, which are similar across the proposed and naive methods. 

\begin{table}[h!]
\centering
\caption{Effect estimates and $95\%$ confidence intervals from the applied example using the national vital statistics dataset from Section \ref{comma-example} for a continuous outcome. Confidence intervals were obtained via the nonparametric bootstrap for the ``EM'', ``PVW'', and ``OLS'' methods. ``EM'', ``PVW'', and ``OLS'' refer to results from the EM algorithm, predictive value weighting, and ordinary least squares correction procedures, respectively, available in the \textbf{COMMA} R package. The ``Naive Analysis'' results are obtained from a standard mediation analysis that does not account for potential misclassification of the mediator variable.} \label{comma-bw-effect-table}
\begin{threeparttable}
\begin{tabular}{l rr rr rr}
\hline
           &
           \multicolumn{2}{c}{\textit{NIE}\tnote{1}}       &
        \multicolumn{2}{c}{\textit{NDE}\tnote{2}}       &
        \multicolumn{2}{c}{\textit{CDE}\tnote{3}} \\
\cline{2-7}
 & \multicolumn{1}{l}{Est.} & \multicolumn{1}{r}{95\% CI} & \multicolumn{1}{l}{Est.} & \multicolumn{1}{r}{95\% CI} & \multicolumn{1}{l}{Est.} & \multicolumn{1}{r}{95\% CI}\\
\hline
\\
EM & 0.014 & (-0.074, 0.078) & -0.095 & (-0.166, -0.005) & -0.090 & (-0.173, -0.012) \\
PVW & -0.008 & (-0.045, 0.016) & -0.078 &  (-0.116, -0.030) & -0.074 & (-0.153, -0.003) \\
OLS &  -0.008 & (-0.017, 0.007) & -0.118 & (-0.405, -0.347) & -0.118 & (-0.405, -0.347) \\
Naive Analysis & 0.007 & (-0.020, 0.034) & -0.090 & (-2.013, 1.834) & -0.088 & (-0.106, -0.070) \\
    \\
\hline  
\end{tabular}
\begin{tablenotes}
\item[1] \textit{NIE} refers to the natural direct effect, which estimates the expected change in birthweight when maternal ethnicity is fixed as non-Hispanic white maternal ethnicity and gestational hypertension status changes to the value it would have attained for each individual if maternal ethnicity was Hispanic. 
\item[2] \textit{NDE} refers to the natural direct effect, which estimates the expected change in birthweightfor non-Hispanic white maternal ethnicity vs. Hispanic maternal ethnicity, while fixing gestational hypertension status as negative. 
\item[3] \textit{CDE} refers to the controlled direct effect, which estimates the expected change in birthweight for non-Hispanic white maternal ethnicity vs. Hispanic maternal ethnicity, conditioning on gestational hypertension status.
\end{tablenotes}
\end{threeparttable}
\end{table}

Next, we can use the $\boldsymbol{\gamma}$ parameter estimates and the \texttt{misclassification\_prob} function from \textbf{COMMA} to estimate average misclassification rates for gestational hypertension. We demonstrate this computation using the $\boldsymbol{\gamma}$ parameter estimates from our proposed EM algorithm.

\begin{lstlisting}
R> bw_misclass_probs_EM <- misclassification_prob(
R>                           matrix(bw_EM$Estimates[10:17],
R>                                  ncol = 2),
R>                           bw_z)
R> bw_misclass_probs_EM %>%
R>   left_join(NCHS2022_sample %>% mutate(Subject = 1:n()) %>%
R>               select(Subject),
R>             by = "Subject") %>%
R>   group_by(M, Mstar) %>%
R>   summarise(mean_pistar = mean(Probability)) %>%
R>   ungroup()
R> # A tibble: 4 x 3
R>       M Mstar mean_pistar
R>   <int> <int>       <dbl>
R> 1     1     1      0.182 
R> 2     1     2      0.818 
R> 3     2     1      0.0893
R> 4     2     2      0.911 
\end{lstlisting}

In the resulting table, \texttt{M} indicates the value of the latent variable for true gestational hypertension status and \texttt{Mstar} indicates the observed value of the \texttt{g\_hyp} variable. For both columns, a value of \texttt{1} indicates a positive diagnosis and a value of \texttt{2} indicates a negative diagnosis. We can also compute sensitivity and specificity estimates using parameter estimates from the PVW and OLS methods in a similar manner. All sensitivity and specificity estimates are provided in Table \ref{bw-probability-results-table}. In context, our estimates suggest that gestational hypertension is diagnosed with high or near-perfect specificity, meaning that few patients are diagnosed with gestational hypertension when the condition is not present. In contrast, our sensitivity estimates are quite low, suggesting that only $18\%$ to $33\%$ of patients who truly have gestational hypertension receive a positive diagnosis. 

\begin{table}[h!]
\centering
\caption{Estimated sensitivity and specificity of gestational hypertension diagnosis from the Section \ref{comma-example} example, using a continuous outcome variable. ``EM'', ``PVW'', and ``OLS'' estimates were computed using the \textbf{COMMA} R Package. The ``Naive'' results were obtained by running a simple logistic regression model for $M^* \sim X, C$ and a simple linear regression model for $Y | X, C, M^*$, assuming no misclassification in $M^*$. Entries of $\boldsymbol{1}$ indicate that a probability was set at 1 for a given method.} \label{bw-probability-results-table}
\begin{threeparttable}
\begin{tabular}{crrrrr}
\hline
\multicolumn{4}{c}{$P(M^* = 1 | M = 1)$} \\
 \cline{1-4}
EM & PVW & OLS & Naive   \\
\hline
0.182 & 0.328 & 0.328 & $\boldsymbol{1}$ \\
\hline
\multicolumn{4}{c}{$P(M^* = 2 | M = 2)$} \\
 \cline{1-4}
EM & PVW & OLS & Naive   \\
\hline
 0.911 & 0.999 & 0.999 & $\boldsymbol{1}$ \\
\hline 
\end{tabular}
\end{threeparttable}
\end{table}

\begin{figure}[h!]
\begin{center}
\includegraphics[width = 450pt]{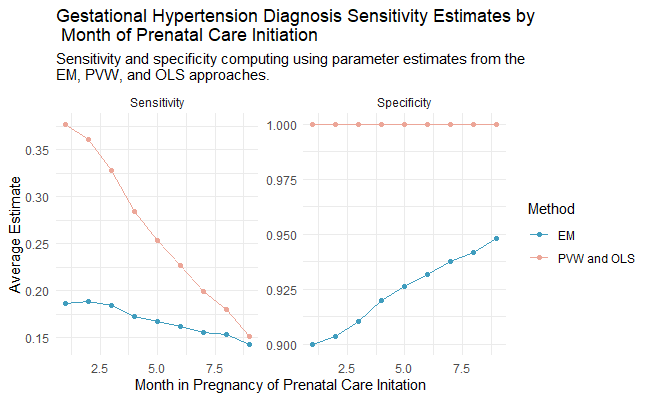}
\caption{Estimated sensitivity and specificity of gestational hypertension diagnosis, computed by month in pregnancy of prenatal care initiation. Sensitivity and specificity estimates were computed using the $\boldsymbol{\gamma}$ parameter estimates from \textbf{COMMA} for the EM algorithm, PVW, and OLS correction approaches.}\label{COMMA_sens_spec_plot}
\end{center}
\end{figure}

Sensitivity and specificity estimates can also be averaged across $\boldsymbol{Z}$ variables. For example, we compute average sensitivity and specificity for each value of \texttt{prenatal\_m} (month of pregnancy in which prenatal care was initiated) in Figure \ref{COMMA_sens_spec_plot}. In this plot, we see that sensitivity decreases as a function of the month of prenatal care initiation for all provided methods. The decrease is steeper when sensitivity is computed using PVW and OLS $\boldsymbol{\gamma}$ parameter estimates than when the quantity is computed using EM $\boldsymbol{\gamma}$ parameter estimates. All provided methods produce parameter estimates that correspond to high specificity estimates, but only the EM algorithm approach shows a positive association between month of prenatal care initiation and gestational hypertension diagnosis specificity. These trends are reasonable in context. The earlier a patient seeks prenatal care, we estimate that the probability that gestational hypertension is correctly diagnosed increases.  

\subsubsection{Parameter estimation with a binary outcome}

In (\ref{eq:COMMA-ex-out-mech}), the outcome variable, \texttt{birth\_wt} is a continuous variable. We also demonstrate COMMA functions for a binary outcome variable by fitting the following model for the \textit{outcome mechanism}:
\begin{equation}
\begin{aligned}
\label{eq:COMMA-ex-out-mech-bin}
\logit\{ P(\textsf{lbw} = 1 | X, \boldsymbol{C}, M; \boldsymbol{\theta}) \} =& \theta_{0} + \theta_{x1} \textsf{ethnicity} + \theta_{c1} \textsf{age\_cs} + \theta_{c2} \textsf{pre\_hyp} + \theta_{c3} \textsf{bmi\_cs} + \\ &\theta_{c4} \textsf{cig\_use} +
\theta_{c5} \textsf{prenatal\_m} + \theta_{c6} \textsf{pre\_diab} + \theta_{c7} \textsf{g\_diab} +\\ & \theta_m M + \theta_{xm} (\textsf{ethnicity} \times M).
\end{aligned}
\end{equation}
In (\ref{eq:COMMA-ex-out-mech-bin}), the outcome variable \texttt{lbw} is a binary variable where a value of $1$ indicates that an infant was classified with a low birth weight (LBW; defined as weight less than $2,500$ grams) and a value of $0$ indicates that the infant was not of low birth weight. 

In order to estimate these models, we first construct the \texttt{lbw} variable by dichotomizing the continuous \texttt{birth\_wt} variable.

\begin{lstlisting}
R> NCHS2022_sample$lbw <- ifelse(NCHS2022_sample$birth_wt < 2.5, 1, 0)
R> lbw_y <- NCHS2022_sample$lbw
\end{lstlisting}

We use obtain $\boldsymbol{\beta}$ and $\boldsymbol{\gamma}$ starting parameters using the same approach as the previous example. We obtain starting values for $\boldsymbol{\theta}$ parameters by fitting a generalized linear model for the \textit{outcome mechanism} that assumes that $M^*$ was measured without error. 

\begin{lstlisting}
R> lbw_beta_glm <- glm(bw_mstar_01 ~ bw_x + bw_c, 
R>                     family = "binomial"(link = "logit"))
R> lbw_beta_start <- matrix(unname(coef(bw_beta_glm)), ncol = 1)
R> lbw_gamma_start <- matrix(rep(0, 8), ncol = 2)
R> lbw_theta_glm <- glm(lbw_y ~ bw_x + bw_mstar_01 + bw_c + 
R>                        bw_x:bw_mstar_01,
R>                      family = "binomial"(link = "logit"))
R> lbw_theta_start <- matrix(unname(coef(lbw_theta_glm)), ncol = 1)
\end{lstlisting}

Next, we use the \texttt{COMMA\_EM} function to estimate model parameters using our proposed EM algorithm. Note that we now specify the outcome distribution as \texttt{"Bernoulli"} since we have dichotomized our outcome variable, \texttt{lbw}. 
\begin{lstlisting}
R> lbw_EM <- COMMA_EM(Mstar = bw_mstar[,1],
R>                    outcome = lbw_y,
R>                    outcome_distribution = "Bernoulli",
R>                    interaction_indicator = TRUE,
R>                    x_matrix = bw_x,
R>                    z_matrix = bw_z,
R>                    c_matrix = bw_c,
R>                    beta_start = lbw_beta_start,
R>                    gamma_start = lbw_gamma_start,
R>                    theta_start = lbw_theta_start,
R>                    tolerance = 1e-7,
R>                    max_em_iterations = 1500,
R>                    em_method = "squarem")
R> lbw_EM
R>    Parameter    Estimates Convergence
R> 1     beta_0  -1.47898138        TRUE
R> 2     beta_1  -0.12011078        TRUE
R> 3     beta_2   0.09392922        TRUE
R> 4     beta_3 -16.92276725        TRUE
R> 5     beta_4   0.54645659        TRUE
R> 6     beta_5  -0.07727433        TRUE
R> 7     beta_6   0.10414145        TRUE
R> 8     beta_7   1.20721909        TRUE
R> 9     beta_8   0.45040585        TRUE
R> 10   gamma11   0.76697167        TRUE
R> 11   gamma21  -0.41009734        TRUE
R> 12   gamma31  -1.09670790        TRUE
R> 13   gamma41  -0.19098768        TRUE
R> 14   gamma12 -83.18493909        TRUE
R> 15   gamma22  -0.07512363        TRUE
R> 16   gamma32 -18.30395513        TRUE
R> 17   gamma42   7.24983487        TRUE
R> 18   theta_0  -3.47009662        TRUE
R> 19   theta_x   0.24390898        TRUE
R> 20   theta_m   1.41516425        TRUE
R> 21  theta_c1   0.04551298        TRUE
R> 22  theta_c2   1.87725224        TRUE
R> 23  theta_c3  -0.15312954        TRUE
R> 24  theta_c4   1.06215624        TRUE
R> 25  theta_c5   0.02895683        TRUE
R> 26  theta_c6   0.53002855        TRUE
R> 27  theta_c7  -0.07738086        TRUE
R> 28  theta_xm   0.22370905        TRUE
\end{lstlisting}

As before, we use the \texttt{COMMA\_EM\_bootstrap\_SE} function to obtain bootstrap standard errors. Only 100 samples are generated to maintain reasonable computation time for this example. 

\begin{lstlisting}
R> lbw_EM_SE <- COMMA_EM_bootstrap_SE(parameter_estimates = 
R>                                       lbw_EM$Estimates,
R>                                    sigma_estimate = NULL,
R>                                    n_bootstrap = 100,
R>                                    n_parallel = 8,
R>                                    outcome_distribution = 
R>                                       "Bernoulli",
R>                                    interaction_indicator = TRUE,
R>                                    x_matrix = bw_x,
R>                                    z_matrix = bw_z,
R>                                    c_matrix = bw_c,
R>                                    tolerance = 1e-7,
R>                                    max_em_iterations = 1500,
R>                                    em_method = "squarem")
R> print(lbw_EM_SE$bootstrap_SE[,c("Parameter", "SE")], n = 28)
R> # A tibble: 28 x 2
R>    Parameter      SE
R>    <chr>       <dbl>
R>  1 beta_0     0.214 
R>  2 beta_1     0.199 
R>  3 beta_2     0.0280
R>  4 beta_3     0.275 
R>  5 beta_4     0.0453
R>  6 beta_5     0.156 
R>  7 beta_6     0.0507
R>  8 beta_7     0.407 
R>  9 beta_8     0.101 
R> 10 gamma11    0.407 
R> 11 gamma12   33.8   
R> 12 gamma21    0.286 
R> 13 gamma22   12.1   
R> 14 gamma31    0.414 
R> 15 gamma32   11.9   
R> 16 gamma41    0.0638
R> 17 gamma42    3.29  
R> 18 theta_0    0.145 
R> 19 theta_c1   0.0313
R> 20 theta_c2   0.206 
R> 21 theta_c3   0.0376
R> 22 theta_c4   0.131 
R> 23 theta_c5   0.0257
R> 24 theta_c6   0.212 
R> 25 theta_c7   0.122 
R> 26 theta_m    0.174 
R> 27 theta_x    0.147 
R> 28 theta_xm   0.239
\end{lstlisting}

Next, we fit our model using the proposed PVW procedure using the \texttt{COMMA\_PVW} function and obtain standard errors with the \texttt{COMMA\_PVW\_bootstrap\_SE} function.

\begin{lstlisting}
R> lbw_PVW <- COMMA_PVW(Mstar = bw_mstar[,1],
R>                      outcome = lbw_y,
R>                      outcome_distribution = "Bernoulli",
R>                      interaction_indicator = TRUE,
R>                      x_matrix = bw_x,
R>                      z_matrix = bw_z,
R>                      c_matrix = bw_c,
R>                      beta_start = bw_beta_start,
R>                      gamma_start = bw_gamma_start,
R>                      theta_start = bw_theta_start,
R>                      tolerance = 1e-7,
R>                      max_em_iterations = 1500,
R>                      em_method = "squarem")
R> lbw_PVW
R>    Parameter    Estimates Convergence Method
R> 1     beta_1  -0.84681217        TRUE    PVW
R> 2     beta_2   0.10776259        TRUE    PVW
R> 3     beta_3   0.10783781        TRUE    PVW
R> 4     beta_4 -18.24508318        TRUE    PVW
R> 5     beta_5   0.77392765        TRUE    PVW
R> 6     beta_6  -0.05728342        TRUE    PVW
R> 7     beta_7   0.11104864        TRUE    PVW
R> 8     beta_8   2.45815410        TRUE    PVW
R> 9     beta_9   0.62272276        TRUE    PVW
R> 10   gamma11  -0.18042109        TRUE    PVW
R> 11   gamma21  -0.50846821        TRUE    PVW
R> 12   gamma31  -0.91513997        TRUE    PVW
R> 13   gamma41  -0.13606180        TRUE    PVW
R> 14   gamma12 -48.90758071        TRUE    PVW
R> 15   gamma22   0.76597815        TRUE    PVW
R> 16   gamma32 -16.88719001        TRUE    PVW
R> 17   gamma42   4.15800323        TRUE    PVW
R> 18   theta_0  -4.05038006        TRUE    PVW
R> 19  theta_x1  -0.17833786        TRUE    PVW
R> 20   theta_m   1.92335466        TRUE    PVW
R> 21  theta_c1   0.03024480        TRUE    PVW
R> 22  theta_c2   2.60033063        TRUE    PVW
R> 23  theta_c3  -0.22703226        TRUE    PVW
R> 24  theta_c4   1.13754667        TRUE    PVW
R> 25  theta_c5   0.03681749        TRUE    PVW
R> 26  theta_c6   0.41425143        TRUE    PVW
R> 27  theta_c7  -0.18571285        TRUE    PVW
R> 28  theta_xm   0.64205976        TRUE    PVW
R> lbw_PVW_SE <- COMMA_PVW_bootstrap_SE(parameter_estimates = 
R>                                         bw_PVW$Estimates,
R>                                      n_bootstrap = 100,
R>                                      n_parallel = 8,
R>                                      outcome_distribution = 
R>                                         "Bernoulli",
R>                                      interaction_indicator = TRUE,
R>                                      x_matrix = bw_x,
R>                                      z_matrix = bw_z,
R>                                      c_matrix = bw_c,
R>                                      tolerance = 1e-7,
R>                                      max_em_iterations = 1500,
R>                                      em_method = "squarem")
R> 
R> print(lbw_PVW_SE$bootstrap_SE[,c("Parameter", "SE")], n = 28)
R> # A tibble: 28 x 2
R>    Parameter      SE
R>    <chr>       <dbl>
R>  1 beta_1     0.231 
R>  2 beta_2     0.221 
R>  3 beta_3     0.0359
R>  4 beta_4     4.53  
R>  5 beta_5     0.0800
R>  6 beta_6     0.185 
R>  7 beta_7     0.0650
R>  8 beta_8     4.80  
R>  9 beta_9     0.180 
R> 10 gamma11    0.230 
R> 11 gamma12   35.9   
R> 12 gamma21    0.166 
R> 13 gamma22    9.85  
R> 14 gamma31    0.331 
R> 15 gamma32   11.1   
R> 16 gamma41    0.0462
R> 17 gamma42    3.37  
R> 18 theta_0    0.255 
R> 19 theta_c1   0.0320
R> 20 theta_c2   0.300 
R> 21 theta_c3   0.0383
R> 22 theta_c4   0.151 
R> 23 theta_c5   0.0284
R> 24 theta_c6   0.250 
R> 25 theta_c7   0.109 
R> 26 theta_m    0.292 
R> 27 theta_x1   0.424 
R> 28 theta_xm   0.489
\end{lstlisting}

We do not compute estimates using the OLS correction method and corresponding \texttt{COMMA\_OLS} function because this method is only appropriate for continuous outcomes. 

We compute effect estimates using the estimated $(\boldsymbol{\beta}, \boldsymbol{\gamma})$ parameters from the EM, PVW, and naive methods in Table \ref{comma-lbw-effect-table} \citep{valeri2013mediation}.

As in the previous example, this analysis is chiefly for demonstration purposes. Larger numbers of bootstrap samples are required to draw rigorous conclusions from the analysis. Preliminarily, our results suggest that accounting for misclassification in gestational hypertension may not meaningfully change effect estimates compared to a naive approach (Table \ref{comma-lbw-effect-table}). A possible exception to this trend is the controlled direct effect (CDE) estimate; under the naive approach, Hispanic maternal ethnicity appears to be associated with increased odds of low birthweight, but this association is non-significant when misdiagnosis of gestational hypertension is accounted for in the PVW and EM algorithm procedures. 

\begin{table}[h!]
\centering
\caption{Effect estimates and $95\%$ confidence intervals from the applied example using the national vital statistics dataset from Section \ref{comma-example} with a binary outcome. Confidence intervals were obtained via the nonparametric bootstrap for the ``EM'' and ``PVW''. ``EM'' and ``PVW''refer to results from the EM algorithm and predictive value weighting procedures, respectively, available in the \textbf{COMMA} R package. The ``Naive Analysis'' results are obtained from a standard mediation analysis that does not account for potential misclassification of the mediator variable.} \label{comma-lbw-effect-table}
\begin{threeparttable}
\begin{tabular}{l rr rr rr}
\hline 
&
           \multicolumn{2}{c}{$OR^{NIE}$\tnote{1}}       &
        \multicolumn{2}{c}{$OR^{NDE}$\tnote{2}}       &
        \multicolumn{2}{c}{$OR^{CDE}$\tnote{3}} \\
        \cline{2-7}
 & \multicolumn{1}{l}{Est.} & \multicolumn{1}{r}{95\% CI} & \multicolumn{1}{l}{Est.} & \multicolumn{1}{r}{95\% CI} & \multicolumn{1}{l}{Est.} & \multicolumn{1}{r}{95\% CI}\\
\hline
\\
EM & 0.959 & (0.855, 1.101) & 1.431 & (1.205, 1.666) & 1.276 & (0.949, 1.616) \\
PVW & 1.060 & (0.892, 1.284) & 1.399 & (1.109, 1.670) & 0.837 & (0.414, 1.892) \\
Naive Analysis & 0.947 & (0.790, 1.134) & 1.453 & (1.181, 1.788) & 1.388 & (1.200, 1.606) \\
    \\
\hline  
\end{tabular}
\begin{tablenotes}
\item[1] $OR^{NIE}$ refers to the natural direct effect, which estimates the expected change in odds of low birthweight when maternal ethnicity is fixed as non-Hispanic white maternal ethnicity and gestational hypertension status changes to the value it would have attained for each individual if maternal ethnicity was Hispanic. 
\item[2] $OR^{NDE}$ refers to the natural direct effect, which estimates the expected in the odds of low birthweight for non-Hispanic white maternal ethnicity vs. Hispanic maternal ethnicity, while fixing gestational hypertension status as negative. 
\item[3] $OR^{CDE}$ refers to the controlled direct effect, which expected change in odds of low birthweight for non-Hispanic white maternal ethnicity vs. Hispanic maternal ethnicity, conditioning on gestational hypertension status.
\end{tablenotes}
\end{threeparttable}
\end{table}

The \texttt{misclassification\_prob} function from \textbf{COMMA} can be used, with the $\boldsymbol{\gamma}$ estimates from the EM algorithm, to compute misclassification rates for the gestational hypertension variable. 

\begin{lstlisting}
R> lbw_misclass_probs_EM <- misclassification_prob(
R>                             matrix(lbw_EM$Estimates[10:17], ncol = 2),
R>                             bw_z)
R> lbw_misclass_probs_EM %>%
R>   left_join(NCHS2022_sample %>% mutate(Subject = 1:n()) %>%
R>               select(Subject),
R>             by = "Subject") %>%
R>   group_by(M, Mstar) %>%
R>   summarise(mean_pistar = mean(Probability)) %>%
R>   ungroup()
R> # A tibble: 4 x 3
R>       M Mstar mean_pistar
R>   <int> <int>       <dbl>
R> 1     1     1    0.517   
R> 2     1     2    0.483   
R> 3     2     1    0.000343
R> 4     2     2    1.00    
\end{lstlisting}

The resulting table is identical in structure to that presented in the birthweight example. As before, we estimate near perfect specificity and imperfect sensitivity in gestational hypertension diagnosis with our proposed EM algorithm. Sensitivity and specificity estimates can be computed similarly using $\boldsymbol{\gamma}$ estimates from the PVW approach. 

We summarize sensitivity and specificity estimates from the EM algorithm, PVW, and naive approaches by maternal ethnicity in Table \ref{lbw-probability-results-table}. As in the previous example, we estimate near-perfect specificity in gestational hypertension diagnosis using the EM algorithm and PVW methods. Our estimates suggest that gestational hypertension diagnosis sensitivity is generally lower for Hispanic mothers compared to non-Hispanic White mothers. This finding would mean that the proportion of Hispanic mothers with gestational hypertension who \textit{are}, in fact, diagnosed with gestational hypertension is lower than that of non-Hispanic White mothers. 

\begin{table}[h!]
\centering
\caption{Estimated sensitivity and specificity of gestational hypertension diagnosis from the Section \ref{comma-example} example, using a binary outcome variable. ``EM'' and ``PVW'' estimates were computed using the \textbf{COMMA} R Package. The ``Naive'' results were obtained by running a simple logistic regression model for $M^* \sim X, C$ and a simple linear regression model for $Y | X, C, M^*$, assuming no misclassification in $M^*$. Entries of $\boldsymbol{1}$ indicate that a probability was set at 1 for a given method.} \label{lbw-probability-results-table}
\begin{threeparttable}
\begin{tabular}{l rrrrr}
\hline
& \multicolumn{3}{c}{$P(M^* = 1 | M = 1)$} \\
 \cline{2-4}
Maternal Ethnicity & EM & PVW & Naive   \\
\hline
Non-Hispanic White & 0.548 & 0.358 & $\boldsymbol{1}$ \\
Hispanic & 0.425 & 0.240 & $\boldsymbol{1}$ \\
\hline
& \multicolumn{3}{c}{$P(M^* = 2 | M = 2)$} \\
 \cline{2-4}
Maternal Ethnicity & EM & PVW& Naive   \\
\hline
Non-Hispanic White & 1.000 & 1.000 & $\boldsymbol{1}$ \\
Hispanic & 1.000 & 0.999 & $\boldsymbol{1}$ \\
\hline 
\end{tabular}
\end{threeparttable}
\end{table}

\section[Conclusion]{Conclusion}\label{discussion}

Binary outcome and mediator variable misclassification can cause bias in parameter estimation in association studies \citep{magder1997logistic, neuhaus1999bias, beesley2020statistical}. Webb and Wells (2023)\cite{webb2023statistical} proposed methods to correct for binary outcome misclassification in cases without a gold standard, without requiring perfect sensitivity or specificity assumptions. Webb, Riley, and Wells (2023)\cite{webb2023assessment} extended these methods to account for potentially misclassified sequential and dependent binary outcome variables using a likelihood-based approach. Multiple methods for conducting a mediation analysis in the presence of a misclassified binary mediator variable are provided in Webb and Wells (2024)\cite{webb2024effect}. 

This paper introduces the R package \textbf{COMBO}, which implements the methods proposed in Webb and Wells (2023)\cite{webb2023statistical} and Webb, Riley, and Wells (2023)\cite{webb2023assessment}. The \textbf{COMBO} R package provides functions to implement expectation-maximization (EM) algorithm and Markov Chain Monte Carlo (MCMC) approaches in cases with one or two potentially misclassified binary outcomes. Other functions in \textbf{COMBO} allow for easy estimation and summarization of estimated sensitivity and specificity rates for the misclassified variable(s).

We also demonstrate the R package \textbf{COMMA} to implement methods from Webb and Wells (2024)\cite{webb2024effect}. \textbf{COMMA} includes functions to an implement an ordinary least squares correction procedure, a predictive value weighting (PVW) approach, and an EM algorithm. The PVW and EM functions in \textbf{COMMA} enable the user to specify the outcome variable distribution and whether or not to include an exposure-mediator interaction term. In addition, \textbf{COMMA} enables users to estimate misclassification rates of the mediator variable.

These user-friendly R packages will allow both statisticians and applied researchers to implement methodologies to correct for misclassification in a variety of analyses. 

\bibliographystyle{ama}
\bibliography{references}

\begin{thebibliography}{10}

\bibitem{omalley2005measuring}
O'{M}alley Kimberly~J, Cook Karon~F, Price Matt~D, Wildes Kimberly~Raiford, Hurdle John~F, Ashton Carol~M. Measuring diagnoses: {ICD} code accuracy  {\it Health {S}ervices {R}esearch. } 2005;40:1620--1639.

\bibitem{ziegler2020binary}
Ziegler Gabriel. Binary Classification Tests, Imperfect Standards, and Ambiguous Information  {\it arXiv preprint arXiv:2012.11215. } 2020.

\bibitem{nguimkeu2021regression}
Nguimkeu Pierre, Rosenman Robert, Tennekoon Vidhura. Regression with a misclassified binary regressor: Correcting for the hidden bias   2021.

\bibitem{grace2016statistical}
Yi~Grace~Y. {\it Statistical {A}nalysis with {M}easurement {E}rror or {M}isclassification}.
\newblock Springer, New York 2016.

\bibitem{beesley2020statistical}
Beesley Lauren~J, Mukherjee Bhramar. Statistical inference for association studies using electronic health records: handling both selection bias and outcome misclassification  {\it Biometrics. } 2022;78:214--226.

\bibitem{zhang2020genetic}
Zhang Qihuang, Yi~Grace~Y. Genetic association studies with bivariate mixed responses subject to measurement error and misclassification  {\it Statistics in Medicine. } 2020;39:3700--3719.

\bibitem{neuhaus1999bias}
Neuhaus John~M. Bias and efficiency loss due to misclassified responses in binary regression  {\it Biometrika. } 1999;86:843--855.

\bibitem{lotspeich2021using}
Lotspeich Sarah~Camilla, others . {\it Using observational data in healthcare research: New methods to design, conduct, and analyze efficient two-phase designs}.
\newblock PhD thesisVanderbilt University 2021.

\bibitem{lotspeich2022efficient}
Lotspeich Sarah~C, Shepherd Bryan~E, Amorim Gustavo~GC, Shaw Pamela~A, Tao Ran. Efficient odds ratio estimation under two-phase sampling using error-prone data from a multi-national HIV research cohort  {\it Biometrics. } 2022;78:1674--1685.

\bibitem{tang2015binary}
Tang Li, Lyles Robert~H., King Caroline~C., Celentano David~D., Lo~Yungtai. Binary regression with differentially misclassified response and exposure variables  {\it Statistics in Medicine. } 2015;34:1605-1620.

\bibitem{lyles2010sensitivity}
Lyles Robert~H, Lin Ji. Sensitivity analysis for misclassification in logistic regression via likelihood methods and predictive value weighting  {\it Statistics in {M}edicine. } 2010;29:2297--2309.

\bibitem{magder1997logistic}
Magder Laurence~S, Hughes James~P. Logistic regression when the outcome is measured with uncertainty  {\it American Journal of Epidemiology. } 1997;146:195--203.

\bibitem{valeri2014estimation}
Valeri Linda, Vanderweele Tyler~J. The estimation of direct and indirect causal effects in the presence of misclassified binary mediator  {\it Biostatistics. } 2014;15:498--512.

\bibitem{webb2024effect}
Webb Kimberly A~Hochstedler, Wells Martin~T. Effect estimation in the presence of a misclassified binary mediator  {\it arXiv preprint arxiv:2407.06970. } 2024.

\bibitem{webb2023statistical}
Webb Kimberly A~Hochstedler, Wells Martin~T. Statistical inference for association studies in the presence of binary outcome misclassification  {\it arXiv preprint arXiv:2303.10215. } 2023.

\bibitem{webb2023assessment}
Webb Kimberly A~Hochstedler, Riley Sarah~A, Wells Martin~T. An assessment of racial disparities in pretrial decision-making using misclassification models  {\it arXiv preprint arXiv:2309.08599. } 2023.

\bibitem{zawistowski2017corrected}
Zawistowski Matthew, Sussman Jeremy~B, Hofer Timothy~P, Bentley Douglas, Hayward Rodney~A, Wiitala Wyndy~L. Corrected ROC analysis for misclassified binary outcomes  {\it Statistics in Medicine. } 2017;36:2148--2160.

\bibitem{SAMBA}
Beesley Lauren. {\it SAMBA: Selection and Misclassification Bias Adjustment for Logistic Regression Models} 2020.
\newblock R package version 0.9.0.

\bibitem{wightman1998lsac}
Wightman Linda~F. LSAC National Longitudinal Bar Passage Study  {\it LSAC Research Report Series. } 1998.

\bibitem{li_direct_2020}
Li~Yige, VanderWeele Tyler~J.. Direct {Effects} under {Differential} {Misclassification} in {Outcomes}, {Exposures}, and {Mediators}  {\it Journal of Causal Inference. } 2020;8:286--299.
\newblock Publisher: De Gruyter.

\bibitem{redner1984mixture}
Redner Richard~A, Walker Homer~F. Mixture densities, maximum likelihood and the {EM} algorithm  {\it SIAM Review. } 1984;26:195--239.

\bibitem{BERRAR2019546}
Berrar Daniel. Performance Measures for Binary Classification  in {\it Encyclopedia of Bioinformatics and Computational Biology} (Ranganathan Shoba, Gribskov Michael, Nakai Kenta, Schönbach Christian. , eds.):546-560Oxford: Academic Press 2019.

\bibitem{collins2014estimation}
Collins John, Huynh Minh. Estimation of diagnostic test accuracy without full verification: a review of latent class methods  {\it Statistics in Medicine. } 2014;33:4141--4169.

\bibitem{duan2021global}
Duan Rui, Ning Yang, Shi Jiasheng, Carroll Raymond~J, Cai Tianxi, Chen Yong. On the global identifiability of logistic regression models with misclassified outcomes  {\it arXiv preprint arXiv:2103.12846. } 2021.

\bibitem{lamont2016regression}
Lamont Andrea~E, Vermunt Jeroen~K, Van~Horn M~Lee. Regression mixture models: Does modeling the covariance between independent variables and latent classes improve the results?  {\it Multivariate Behavioral Research. } 2016;51:35--52.

\bibitem{jones2010identifiability}
Jones Geoffrey, Johnson Wesley~O, Hanson Timothy~E, Christensen Ronald. Identifiability of models for multiple diagnostic testing in the absence of a gold standard  {\it Biometrics. } 2010;66:855--863.

\bibitem{dempster1977maximum}
Dempster Arthur~P, Laird Nan~M, Rubin Donald~B. Maximum likelihood from incomplete data via the {EM} algorithm  {\it Journal of the Royal Statistical Society: Series B (Methodological). } 1977;39:1--22.

\bibitem{agresti2003categorical}
Agresti Alan. {\it Categorical Data Analysis}.
\newblock New York: John Wiley \& Sons 2003.

\bibitem{dplyr}
Wickham Hadley, François Romain, Henry Lionel, Müller Kirill, Vaughan Davis. {\it dplyr: A Grammar of Data Manipulation} 2023.
\newblock R package version 1.1.4, https://github.com/tidyverse/dplyr.

\bibitem{lovins2016validation}
Lovins Brian, Lovins Lori. Validation of a pretrial risk assessment tool  {\it Correctional Consultants. } 2016.

\bibitem{cadigan2011implementing}
Cadigan Timothy~P, Lowenkamp Christopher~T. Implementing risk assessment in the federal pretrial services system  {\it Fed. Probation. } 2011;75:30-34.

\bibitem{stevenson2018assessing}
Stevenson Megan. Assessing risk assessment in action  {\it Minn. L. Rev.. } 2018;103:303.

\bibitem{stevenson2022algorithmic}
Stevenson Megan~T, Doleac Jennifer~L. Algorithmic risk assessment in the hands of humans  {\it Available at SSRN 3489440. } 2022.

\bibitem{copp2022pretrial}
Copp Jennifer~E, Casey William, Blomberg Thomas~G, Pesta George. Pretrial risk assessment instruments in practice: The role of judicial discretion in pretrial reform  {\it Criminology \& Public Policy. } 2022;21:329--358.

\bibitem{coston2021characterizing}
Coston Amanda, Rambachan Ashesh, Chouldechova Alexandra. Characterizing fairness over the set of good models under selective labels  in {\it International Conference on Machine Learning}:2144--2155 2021.

\bibitem{fogliato2020fairness}
Fogliato Riccardo, Chouldechova Alexandra, G’Sell Max. Fairness evaluation in presence of biased noisy labels  in {\it International {C}onference on {A}rtificial {I}ntelligence and {S}tatistics}:2325--2336 2020.

\bibitem{lakkaraju2017selective}
Lakkaraju Himabindu, Kleinberg Jon, Leskovec Jure, Ludwig Jens, Mullainathan Sendhil. The selective labels problem: Evaluating algorithmic predictions in the presence of unobservables  in {\it Proceedings of the 23rd ACM SIGKDD International Conference on Knowledge Discovery and Data Mining}:275--284 2017.

\bibitem{NCHS}
NCHS . {\it National Center for Health Statistic}.
\newblock U.S. Department of Health and Human Services 2022.

\bibitem{valeri2013mediation}
Valeri Linda, VanderWeele Tyler~J. Mediation analysis allowing for exposure--mediator interactions and causal interpretation: Theoretical assumptions and implementation with SAS and SPSS macros.  {\it Psychological Methods. } 2013;18:137.

\end{thebibliography}

\begin{appendices}

\section[Evaluating a risk assessment algorithm in COMBO via MCMC]{Evaluating a risk assessment algorithm in COMBO via MCMC}\label{COMBO-MCMC-2stage-appendix}

In this section, we provide code to perform a Markov Chain Monte Carlo (MCMC) analysis for the scenario presented in Section \ref{COMBO-2stage-example}. First, we specify prior distributions for $\boldsymbol{\beta}$, $\boldsymbol{\gamma_1}$, and $\boldsymbol{\gamma_2}$.

\begin{lstlisting}
R> prior_distribution <- "normal"
R> normal_mu_beta <- matrix(c(0, 0, 0, 0, 0, NA, NA, NA, NA, NA),
R>                          nrow = 2, byrow = TRUE)
R> normal_sd_beta <- matrix(c(10, 10, 10, 10, 10, NA, NA, NA, NA, NA),
R>                          nrow = 2, byrow = TRUE)
R> normal_mu_gamma1 <- array(data = c(0, NA, 0, NA, 0, NA, 0, NA),
R>                           dim = c(2,2,2))
R> normal_sd_gamma1 <- array(data = c(10, NA, 10, NA, 10, NA, 10, NA),
R>                           dim = c(2,2,2))
R> normal_mu_gamma2 <- array(rep(c(0, NA), 8), dim = c(2,2,2,2))
R> normal_sd_gamma2 <- array(rep(c(10, NA), 8), dim = c(2,2,2,2))
R> beta_prior_parameters <- list(mu = normal_mu_beta,
R>                               sigma = normal_sd_beta)
R> gamma1_prior_parameters <- list(mu = normal_mu_gamma1,
R>                                 sigma = normal_sd_gamma1)
R> gamma2_prior_parameters <- list(mu = normal_mu_gamma2,
R>                                 sigma = normal_sd_gamma2)
R> normal_mu_naive_gamma2 <- normal_mu_gamma1
R> normal_sd_naive_gamma2 <- normal_sd_gamma1
R> naive_gamma2_prior_parameters <- list(mu = normal_mu_naive_gamma2,
R>                                       sigma = normal_sd_naive_gamma2)
\end{lstlisting}

Using these prior settings, we run the MCMC analysis with the \texttt{COMBO\_MCMC\_2stage} function in \textbf{COMBO}.

\begin{lstlisting}
R> vprai_MCMC <- COMBO_MCMC_2stage(Ystar1 = Ystar1,
R>                                 Ystar2 = Ystar2,
R>                                 x_matrix = x,
R>                                 z1_matrix = z1, 
R>                                 z2_matrix = z2,
R>                                 prior = prior_distribution,
R>                                 beta_prior_parameters,
R>                                 gamma1_prior_parameters,
R>                                 gamma2_prior_parameters,
R>                                 naive_gamma2_prior_parameters,
R>                                 number_MCMC_chains = 2,
R>                                 MCMC_sample = 1000,
R>                                 burn_in = 500)
\end{lstlisting}

Results from the \texttt{COMBO\_MCMC\_2stage} function are saved in a list, which contains posterior means and medians for misclassification model analysis as well as posterior means and medians for a naive approach that does not account for outcome misclassification.
\begin{lstlisting}
R> vprai_MCMC$posterior_means_df
R> # A tibble: 17 x 3
R>    parameter_name  posterior_mean posterior_median
R>    <fct>                    <dbl>            <dbl>
R>  1 beta[1,1]              -2.18            -2.18  
R>  2 beta[1,2]               0.799            0.797 
R>  3 beta[1,3]               0.358            0.374 
R>  4 beta[1,4]               1.21             1.21  
R>  5 beta[1,5]               0.489            0.481 
R>  6 gamma1[1,1,1]          -0.782           -0.801 
R>  7 gamma1[1,2,1]          -3.01            -3.01  
R>  8 gamma1[1,1,2]           0.780            0.770 
R>  9 gamma1[1,2,2]          -0.574           -0.570 
R> 10 gamma2[1,1,1,1]         0.603            0.584 
R> 11 gamma2[1,2,1,1]         0.723            0.725 
R> 12 gamma2[1,1,2,1]        -0.0425          -0.0445
R> 13 gamma2[1,2,2,1]        -0.498           -0.496 
R> 14 gamma2[1,1,1,2]         0.483            0.486 
R> 15 gamma2[1,2,1,2]         0.455            0.453 
R> 16 gamma2[1,1,2,2]         0.0160           0.0146
R> 17 gamma2[1,2,2,2]         0.435            0.436
R> vprai_MCMC$naive_posterior_means_df
R> # A tibble: 9 x 3
R>   parameter_name      posterior_mean posterior_median
R>   <chr>                        <dbl>            <dbl>
R> 1 naive_beta[1,1]            -2.95           -2.95   
R> 2 naive_beta[1,2]             0.819           0.821  
R> 3 naive_beta[1,3]             0.353           0.350  
R> 4 naive_beta[1,4]             1.14            1.14   
R> 5 naive_beta[1,5]             0.221           0.220  
R> 6 naive_gamma2[1,1,1]         0.482           0.413  
R> 7 naive_gamma2[1,1,2]         6.13            6.17   
R> 8 naive_gamma2[1,2,1]         0.0125          0.00868
R> 9 naive_gamma2[1,2,2]         0.0701          0.0599
\end{lstlisting}

\section[Evaluating a risk assessment algorithm with synthetic data in COMBO]{Evaluating a risk assessment algorithm with synthetic data in \textbf{COMBO}} \label{vprai-synthetic-example}

In this section, we replicate the risk assessment algorithm study presented in Section \ref{COMBO-2stage-example} using a synthetic dataset. The synthetic dataset is based off of the original dataset (denoted \texttt{vprai\_data} in Section \ref{COMBO-2stage-example}), but all variables have been perturbed and all potential identifying information is removed. The synthetic dataset is available in \textbf{COMBO} and can be loaded using the following code:
\begin{lstlisting}
R> data("VPRAI_synthetic_data")
\end{lstlisting}

Our goal is to fit the model in Equation (\ref{eq:vprai-example-mechanisms}). First, we create matrices for the predictors and observed outcomes: 
\begin{lstlisting}
R> Ystar1 <- VPRAI_synthetic_data$vprai
R> Ystar1_01 <- ifelse(Ystar1 == 2, 0, 1)
R> Ystar2 <- VPRAI_synthetic_data$judge
R> Ystar2_01 <- ifelse(Ystar2 == 2, 0, 1)
R> x <- matrix(c(VPRAI_synthetic_data$n_FTA,
R>               VPRAI_synthetic_data$unemployed,
R>               VPRAI_synthetic_data$drug_abuse,
R>               VPRAI_synthetic_data$n_violent_arrest),
R>             ncol = 4, byrow = FALSE)
R> z1 <- matrix(c(VPRAI_synthetic_data$race),
R>              ncol = 1, byrow = FALSE)
R> z2 <- matrix(c(VPRAI_synthetic_data$race),
R>              ncol = 1, byrow = FALSE)
\end{lstlisting}

We fit a naive regression model to obtain $\boldsymbol{\beta}$ starting values for the EM algorithm. We set all starting values for the $\boldsymbol{\gamma}$ terms at 0. 

\begin{lstlisting}
R> beta_start_glm <- glm(Ystar1_01 ~ x,
R>                       family = "binomial"(link = "logit"))
R> beta_start <- matrix(c(unname(coef(beta_start_glm))), ncol = 1)
R> gamma1_start <- matrix(rep(0, 4), ncol = 2)
R> gamma2_start <- array(rep(0, 8), dim = c(2,2,2))
\end{lstlisting}

We use the \texttt{COMBO\_EM\_2stage} function to estimate model parameters in Equation (\ref{eq:vprai-example-mechanisms}). This function also performs a naive analysis that ignores misclassification in $Y^{*(1)}$ and $Y^{*(2)}$.

\begin{lstlisting}
R> vprai_EM <- COMBO_EM_2stage(Ystar1 = Ystar1,
R>                             Ystar2 = Ystar2,
R>                             x_matrix = x,
R>                             z1_matrix = z1, 
R>                             z2_matrix = z2,
R>                             beta_start = beta_start,
R>                             gamma1_start = gamma1_start,
R>                             gamma2_start = gamma2_start)
R> vprai_EM
R>          Parameter    Estimates           SE Convergence
R> 1           beta_1  -3.64455465   0.20468008        TRUE
R> 2           beta_2   1.74994558   2.04082886        TRUE
R> 3           beta_3   0.86824584   0.29542410        TRUE
R> 4           beta_4   2.35178957   0.33029463        TRUE
R> 5           beta_5   0.31466301   0.11294888        TRUE
R> 6        gamma1_11  -0.42370291   0.31045651        TRUE
R> 7        gamma1_21   0.87676085   0.60069239        TRUE
R> 8        gamma1_12  -6.05133307   0.60437079        TRUE
R> 9        gamma1_22   2.39742301   0.72241095        TRUE
R> 10     gamma2_1111   1.96103236   0.31601917        TRUE
R> 11     gamma2_2111   0.36105769   0.43231762        TRUE
R> 12     gamma2_1121   0.36036796   0.11673536        TRUE
R> 13     gamma2_2121   0.27894658   0.67230801        TRUE
R> 14     gamma2_1112  16.82634967   2.78006173        TRUE
R> 15     gamma2_2112 -15.87894404   2.77152124        TRUE
R> 16     gamma2_1122  -0.45710716   0.07393581        TRUE
R> 17     gamma2_2122   0.46281381   0.10698544        TRUE
R> 18    naive_beta_1   0.15086322 -22.35192309        TRUE
R> 19    naive_beta_2   0.12419324   5.64411648        TRUE
R> 20    naive_beta_3   0.14923273   2.88803540        TRUE
R> 21    naive_beta_4   0.15394199   9.38240661        TRUE
R> 22    naive_beta_5   0.02982090   6.57863026        TRUE
R> 23 naive_gamma2_11   0.32157058   6.17012708        TRUE
R> 24 naive_gamma2_21   0.40737381   0.21685613        TRUE
R> 25 naive_gamma2_12   0.06714098  -4.94426281        TRUE
R> 26 naive_gamma2_22   0.09673119   4.18307528        TRUE
\end{lstlisting}

We can also estimate the model using an MCMC procedure via the \texttt{COMBO\_MCMC\_2stage} function in \textbf{COMBO}.

\begin{lstlisting}
R> prior_distribution <- "normal"
R> normal_mu_beta <- matrix(c(0, 0, 0, 0, 0, NA, NA, NA, NA, NA),
R>                          nrow = 2, byrow = TRUE)
R> normal_sd_beta <- matrix(c(10, 10, 10, 10, 10, NA, NA, NA, NA, NA),
R>                          nrow = 2, byrow = TRUE)
R> normal_mu_gamma1 <- array(data = c(0, NA, 0, NA, 0, NA, 0, NA),
R>                           dim = c(2,2,2))
R> normal_sd_gamma1 <- array(data = c(10, NA, 10, NA, 10, NA, 10, NA),
R>                           dim = c(2,2,2))
R> normal_mu_gamma2 <- array(rep(c(0, NA), 8), dim = c(2,2,2,2))
R> normal_sd_gamma2 <- array(rep(c(10, NA), 8), dim = c(2,2,2,2))
R> beta_prior_parameters <- list(mu = normal_mu_beta,
R>                               sigma = normal_sd_beta)
R> gamma1_prior_parameters <- list(mu = normal_mu_gamma1,
R>                                 sigma = normal_sd_gamma1)
R> gamma2_prior_parameters <- list(mu = normal_mu_gamma2,
R>                                 sigma = normal_sd_gamma2)
R> normal_mu_naive_gamma2 <- normal_mu_gamma1
R> normal_sd_naive_gamma2 <- normal_sd_gamma1
R> naive_gamma2_prior_parameters <- list(mu = normal_mu_naive_gamma2,
R>                                       sigma = normal_sd_naive_gamma2)
R> vprai_MCMC <- COMBO_MCMC_2stage(Ystar1 = Ystar1,
R>                                 Ystar2 = Ystar2,
R>                                 x_matrix = x,
R>                                 z1_matrix = z1, 
R>                                 z2_matrix = z2,
R>                                 prior = prior_distribution,
R>                                 beta_prior_parameters,
R>                                 gamma1_prior_parameters,
R>                                 gamma2_prior_parameters
R>                                 naive_gamma2_prior_parameters,
R>                                 number_MCMC_chains = 2,
R>                                 MCMC_sample = 1000,
R>                                 burn_in = 500)
R> vprai_MCMC$posterior_means_df
R> # A tibble: 17 x 3
R>    parameter_name  posterior_mean posterior_median
R>    <fct>                    <dbl>            <dbl>
R>  1 beta[1,1]              -2.05            -2.06  
R>  2 beta[1,2]               0.857            0.853 
R>  3 beta[1,3]               0.327            0.314 
R>  4 beta[1,4]               1.23             1.23  
R>  5 beta[1,5]               0.260            0.260 
R>  6 gamma1[1,1,1]          -0.772           -0.790 
R>  7 gamma1[1,2,1]          -2.87            -2.87  
R>  8 gamma1[1,1,2]           0.648            0.642 
R>  9 gamma1[1,2,2]          -0.571           -0.571 
R> 10 gamma2[1,1,1,1]         0.451            0.456 
R> 11 gamma2[1,2,1,1]         0.590            0.580 
R> 12 gamma2[1,1,2,1]        -0.0217          -0.0123
R> 13 gamma2[1,2,2,1]        -0.448           -0.447 
R> 14 gamma2[1,1,1,2]         0.403            0.412 
R> 15 gamma2[1,2,1,2]         0.446            0.442 
R> 16 gamma2[1,1,2,2]         0.0259           0.0410
R> 17 gamma2[1,2,2,2]         0.434            0.434
R> vprai_MCMC$naive_posterior_means_df
R> # A tibble: 9 x 3
R>   parameter_name      posterior_mean posterior_median
R>   <chr>                        <dbl>            <dbl>
R> 1 naive_beta[1,1]            -2.78            -2.78  
R> 2 naive_beta[1,2]             0.593            0.591 
R> 3 naive_beta[1,3]             0.265            0.268 
R> 4 naive_beta[1,4]             1.00             1.00  
R> 5 naive_beta[1,5]             0.157            0.157 
R> 6 naive_gamma2[1,1,1]         0.397            0.315 
R> 7 naive_gamma2[1,1,2]         6.28             6.37  
R> 8 naive_gamma2[1,2,1]         0.0139           0.0103
R> 9 naive_gamma2[1,2,2]         0.0664           0.0559
\end{lstlisting}

Our goal is to evaluate the predictive accuracy of our misclassification model for pretrial failure to VPRAI recommendations. We evaluate these risk assessments through a receiver operating curve (ROC) analysis. For this analysis, we first use the $\boldsymbol{\beta}$ estimates from the EM algorithm to compute bias-corrected pretrial failure probabilities for all observations in the dataset. 

\begin{lstlisting}
R> beta_estimates <- matrix(vprai_EM$Estimates[1:5], ncol = 1)
R> p_pretrial_failure <- true_classification_prob(beta_estimates,
R>                                                x)
R> VPRAI_synthetic_data$p_pretrial_failure <- 
R>    p_pretrial_failure$Probability[which(p_pretrial_failure$Y == 1)]
\end{lstlisting}

Because we do not have ``ground truth'' pretrial failure outcomes for every individual in the dataset, we follow the misclassification-adjusted ROC analysis procedure from \cite{zawistowski2017corrected}. This procedure requires estimates of the conditional predictive probability of pretrial failure for each subject. We estimate these probabilities using an internal \textbf{COMBO} function. 
\begin{lstlisting}
R> Ystar1_matrix <- matrix(c(Ystar1_01, 1 - Ystar1_01),
R>                         ncol = 2, byrow = FALSE)
R> Ystar2_matrix <- matrix(c(Ystar2_01, 1 - Ystar2_01),
R>                         ncol = 2, byrow = FALSE)
R> X_design <- matrix(c(rep(1, nrow(x)), c(x)),
R>                    nrow = nrow(x), byrow = FALSE)
R> pi_matrix <- COMBO:::pi_compute(beta_estimates, X_design, 
R>                                 nrow(X_design), 2)
R> Z1_design <- matrix(c(rep(1, nrow(z1)), c(z1)),
R>                     ncol = 2, byrow = FALSE)
R> gamma1_estimates <- matrix(vprai_EM$Estimates[6:9],
R>                            ncol = 2, byrow = FALSE)
R> pistar1_matrix <- COMBO:::pistar_compute(gamma1_estimates, Z1_design,
R>                                          nrow(Z1_design), 2)
R> Z2_design <- matrix(c(rep(1, nrow(z2)), c(z2)),
R>                     ncol = 2, byrow = FALSE)
R> gamma2_estimates <- array(vprai_EM$Estimates[10:17], dim = c(2,2,2))
R> pistar2_array <- COMBO:::pitilde_compute(gamma2_estimates, Z2_design,
R>                                          nrow(Z2_design), 2)
R> predictive_prob_pf <- COMBO:::w_j_2stage(Ystar1_matrix,
R>                                          Ystar2_matrix,
R>                                          pistar2_array,
R>                                          pistar1_matrix,
R>                                          pi_matrix, 
R>                                          nrow(Ystar1_matrix), 2)
R> VPRAI_synthetic_data$predictive_prob_pf <- predictive_prob_pf[,1]
\end{lstlisting}

The resulting probabilities are used to compute misclassification-adjusted true positive rate (TPR) and false positive rate (FPR) estimates \citep{zawistowski2017corrected}. We use these rates to construct an ROC curve and estimate the area under the curve (AUC). 

\begin{lstlisting}
R> cutoffs <- seq(0, 1, by = .01)
R> bias_corrected_TPR <- rep(NA, length(cutoffs))
R> bias_corrected_FPR <- rep(NA, length(cutoffs))
R> for(i in 1:length(cutoffs)){
R>   
R>   cutoff_i <- cutoffs[i]
R>   model_recommendation <- ifelse(
R>     VPRAI_synthetic_data$p_pretrial_failure > cutoff_i, 1, 0)
R>   
R>   bias_corrected_TPR[i] <- sum(predictive_prob_pf[,1] * 
R>     model_recommendation) /
R>     sum(predictive_prob_pf[,1])
R>   
R>   bias_corrected_FPR[i] <- sum(predictive_prob_pf[,2] *
R>     model_recommendation) /
R>     sum(predictive_prob_pf[,2])  
R> }
R> 
R> bias_corrected_roc_data <- data.frame(TPR = bias_corrected_TPR,
R>                                       FPR = bias_corrected_FPR)
R> 
R> ggplot(data = bias_corrected_roc_data) +
R>   geom_line(aes(x = FPR, y = TPR), color = "#409DBE") +
R>   geom_point(aes(x = FPR, y = TPR), color = "#409DBE") +
R>   theme_minimal() +
R>   geom_abline(slope = 1, intercept = 0) 
R> 
R> trapz(fliplr(bias_corrected_FPR), fliplr(bias_corrected_TPR))
R> [1] 0.866636
\end{lstlisting}

\begin{figure}[h!]
\begin{center}
\includegraphics[width = \textwidth]{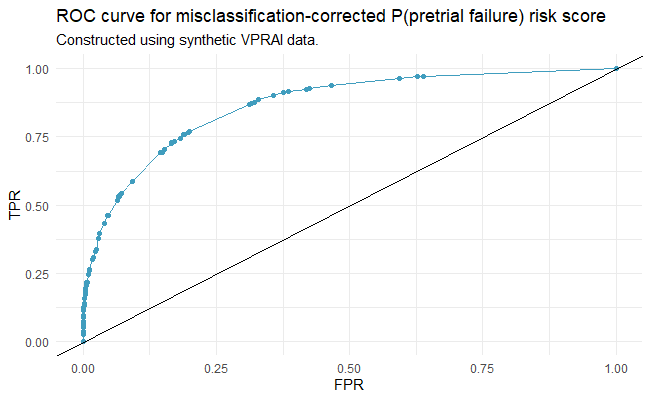}
\caption{Receiver-operating characteristic (ROC) curve for the misclassification-corrected predicted probability of pretrial failure from the synthetic VPRAI pretrial dataset. The ROC curve is obtained using the misclassification-adjusted ROC approach from \cite{zawistowski2017corrected}. FPR stands for false positive rate and TPR stands for true positive rate.}\label{synthetic-VPRAI-ROC}
\end{center}
\end{figure}

The resulting AUC estimate is 0.867. The resulting ROC curve is provided in Figure \ref{synthetic-VPRAI-ROC}.

Next, we evaluate the predictive accuracy of our bias-corrected model compared to VPRAI recommendations only among individuals in the dataset who were released ahead of their trial date. We first define this subset, and then compare our model-estimated probability of pretrial failure among individuals who actually did and did not experience pretrial failure. 

\begin{lstlisting}
R> release_subset <- VPRAI_synthetic_data %>%
R>   filter(judge == 2)
R> release_subset %>%
R>   group_by(pretrial_failure) %>%
R>   summarise(avg_p_pretrial_failure = mean(p_pretrial_failure))
R> # A tibble: 2 x 2
R>   pretrial_failure avg_p_pretrial_failure
R>              <int>                  <dbl>
R> 1                0                  0.169
R> 2                1                  0.453
\end{lstlisting}

In the synthetic dataset, the estimated probability of pretrial failure was 0.166 among individuals who truly did not experience a pretrial failure and was 0.481 among individuals who truly did experience a pretrial failure. 

\begin{figure}[h!]
\begin{center}
\includegraphics[width = \textwidth]{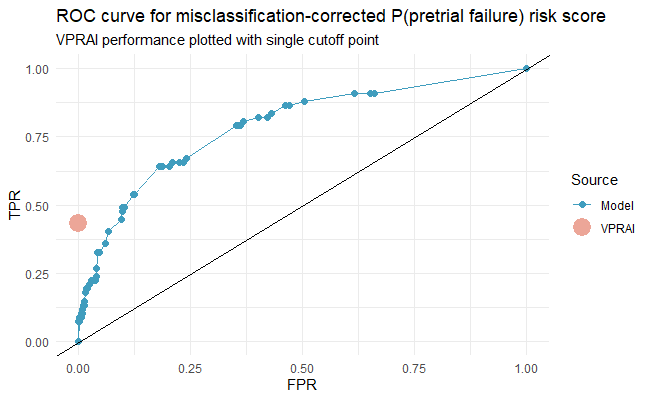}
\caption{Receiver-operating characteristic (ROC) curve for the misclassification-corrected predicted probability of pretrial failure from the synthetic VPRAI pretrial dataset, limited to released individuals who have gold standard pretrial failure outcomes. TPR and FPR of the synthetic VPRAI are included as a single point, since multiple cutoffs cannot be evaluated for the VPRAI algorithm (the VPRAI algorithm returns a ``release'' or ``detain'' recommendation, rather than a continuous risk prediction). FPR stands for false positive rate and TPR stands for true positive rate.}\label{synthetic-subset-ROC}
\end{center}
\end{figure}

Using this subset, we construct an ROC curve (Figure \ref{synthetic-subset-ROC}) and estimate the AUC for our bias-corrected risk estimates. We include the VPRAI accuracy as a single point because it only provides a recommendation output, rather than a range of risk probabilities. 

\begin{lstlisting}
R> cutoffs <- seq(0, 1, by = .01)
R> model_sensitivity <- rep(NA, length(cutoffs))
R> model_specificity <- rep(NA, length(cutoffs))
R> for(i in 1:length(cutoffs)){
R>   
R>   cutoff_i <- cutoffs[i]
R>   model_recommendation <- ifelse(
R>     release_subset$p_pretrial_failure > cutoff_i, 1, 0)
R>   
R>   model_sensitivity[i] <- (length(which(model_recommendation == 1 &
R>     release_subset$pretrial_failure == 1))) /
R>     length(which(release_subset$pretrial_failure == 1))
R>   
R>   model_specificity[i] <- (length(which(model_recommendation == 0 &
R>     release_subset$pretrial_failure == 0))) /
R>     length(which(release_subset$pretrial_failure == 0)) 
R> }
R> 
R> VPRAI_sensitivity <- (length(which(release_subset$vprai == 1 &
R>                         release_subset$pretrial_failure == 1))) /
R>   length(which(release_subset$pretrial_failure == 1))
R> 
R> VPRAI_specificity <- (length(which(release_subset$vprai == 2 &
R>                         release_subset$pretrial_failure == 0))) /
R>   length(which(release_subset$pretrial_failure == 0))
R> 
R> model_roc_data <- data.frame("FPR" = c(1 - model_specificity,
R>                                        1 - VPRAI_specificity),
R>                              "TPR" = c(model_sensitivity,
R>                                        VPRAI_sensitivity),
R>                              "Source" = c(rep("Model", 101),
R>                                           "VPRAI"))
R> 
R> ggplot(data = model_roc_data) +
R>   geom_line(aes(x = FPR, y = TPR, color = Source)) +
R>   geom_point(aes(x = FPR, y = TPR, color = Source, size = Source)) +
R>   theme_minimal() +
R>   geom_abline(slope = 1, intercept = 0) +
R>   scale_color_manual(values = c("#409DBE", "#ECA698"))
R> 
R> ## Model AUC
R> trapz(fliplr(1 - model_specificity), fliplr(model_sensitivity))
R> [1] 0.789579
R> ## VPRAI AUC
R> release_subset$vprai_01 <- ifelse(release_subset$vprai == 1, 1, 0)
R> auc(release_subset$pretrial_failure, release_subset$vprai_01)
R> Setting levels: control = 0, case = 1
R> Setting direction: controls < cases
R> Area under the curve: 0.7164
\end{lstlisting}

In this synthetic example, our bias-corrected risk predictions achieve a higher AUC than that of the VPRAI, 0.790 vs. 0.716, respectively. 

\section{PVW and OLS Estimation Functions}\label{COMMA-PVS-OLS-appendix}
The \textbf{COMMA} R package supports parameter estimation via the predictive value weighting (PVW) approach and ordinary least squares (OLS) correction from Section \ref{comma-pvw} and Section \ref{comma-ols}, respectively. The \texttt{COMMA\_PVW} function uses the PVW approach to estimate $(\boldsymbol{\beta}, \boldsymbol{\gamma}, \boldsymbol{\theta})$, demonstrated below.
\begin{lstlisting}
R> bw_PVW <- COMMA_PVW(Mstar = bw_mstar[,1],
R>                     outcome = bw_y,
R>                     outcome_distribution = "Normal",
R>                     interaction_indicator = TRUE,
R>                     x_matrix = bw_x,
R>                     z_matrix = bw_z,
R>                     c_matrix = bw_c,
R>                     beta_start = bw_beta_start,
R>                     gamma_start = bw_gamma_start,
R>                     theta_start = bw_theta_start,
R>                     tolerance = 1e-7,
R>                     max_em_iterations = 1500,
R>                     em_method = "squarem")
R> bw_PVW
R>    Parameter     Estimates Convergence Method
R> 1     beta_1  -0.846812169        TRUE    PVW
R> 2     beta_2   0.107762594        TRUE    PVW
R> 3     beta_3   0.107837807        TRUE    PVW
R> 4     beta_4 -18.245083179        TRUE    PVW
R> 5     beta_5   0.773927649        TRUE    PVW
R> 6     beta_6  -0.057283420        TRUE    PVW
R> 7     beta_7   0.111048637        TRUE    PVW
R> 8     beta_8   2.458154099        TRUE    PVW
R> 9     beta_9   0.622722764        TRUE    PVW
R> 10   gamma11  -0.180421091        TRUE    PVW
R> 11   gamma21  -0.508468206        TRUE    PVW
R> 12   gamma31  -0.915139970        TRUE    PVW
R> 13   gamma41  -0.136061801        TRUE    PVW
R> 14   gamma12 -48.907580707        TRUE    PVW
R> 15   gamma22   0.765978153        TRUE    PVW
R> 16   gamma32 -16.887190012        TRUE    PVW
R> 17   gamma42   4.158003235        TRUE    PVW
R> 18   theta_0   3.450996520        TRUE    PVW
R> 19  theta_x1  -0.074268269        TRUE    PVW
R> 20   theta_m  -0.318492971        TRUE    PVW
R> 21  theta_c1   0.013010721        TRUE    PVW
R> 22  theta_c2  -0.479736529        TRUE    PVW
R> 23  theta_c3   0.080919260        TRUE    PVW
R> 24  theta_c4  -0.240684961        TRUE    PVW
R> 25  theta_c5  -0.007136029        TRUE    PVW
R> 26  theta_c6  -0.025735289        TRUE    PVW
R> 27  theta_c7   0.020213785        TRUE    PVW
R> 28  theta_xm  -0.011172444        TRUE    PVW
\end{lstlisting}

The \texttt{COMMA\_PVW\_bootstrap\_SE} function allows us to compute bootstrap standard error estimates. 

\begin{lstlisting}
R> bw_PVW_SE <- COMMA_PVW_bootstrap_SE(parameter_estimates = 
R>                                       bw_PVW$Estimates,
R>                                     sigma_estimate = 1,
R>                                     n_bootstrap = 100,
R>                                     n_parallel = 8,
R>                                     outcome_distribution = "Normal",
R>                                     interaction_indicator = TRUE,
R>                                     x_matrix = bw_x,
R>                                     z_matrix = bw_z,
R>                                     c_matrix = bw_c,
R>                                     tolerance = 1e-7,
R>                                     max_em_iterations = 1500,
R>                                     em_method = "squarem")
R> print(bw_PVW_SE$bootstrap_SE[,c("Parameter", "SE")], n = 28)
R> # A tibble: 28 x 2
R>    Parameter       SE
R>    <chr>        <dbl>
R>  1 beta_1     0.239  
R>  2 beta_2     0.246  
R>  3 beta_3     0.0329 
R>  4 beta_4     4.57   
R>  5 beta_5     0.0682 
R>  6 beta_6     0.212  
R>  7 beta_7     0.0724 
R>  8 beta_8     4.81   
R>  9 beta_9     0.177  
R> 10 gamma11    0.205  
R> 11 gamma12   37.0    
R> 12 gamma21    0.177  
R> 13 gamma22    9.46   
R> 14 gamma31    0.331  
R> 15 gamma32   11.3    
R> 16 gamma41    0.0510 
R> 17 gamma42    3.29   
R> 18 theta_0    0.0284 
R> 19 theta_c1   0.00771
R> 20 theta_c2   0.0560 
R> 21 theta_c3   0.0100 
R> 22 theta_c4   0.0483 
R> 23 theta_c5   0.00627
R> 24 theta_c6   0.0793 
R> 25 theta_c7   0.0344 
R> 26 theta_m    0.0401 
R> 27 theta_x1   0.0403 
R> 28 theta_xm   0.0778
\end{lstlisting}

The \texttt{COMMA\_OLS} function can be used to estimate $(\boldsymbol{\beta}, \boldsymbol{\gamma}, \boldsymbol{\theta})$ with the OLS correction approach. Note that the OLS correction only applies to Normally distributed outcome variables, so we do not have a function argument to indicate the outcome distribution. In addition, the OLS correction method does not support inclusion of an interaction term between $X$ and $M$ in the \textit{outcome mechanism}, so an argument indicating whether an interaction term should be estimated is not required in \texttt{COMMA\_OLS}. This also means that the last value in \texttt{bw\_theta\_start} is omitted in the \texttt{theta\_start} argument, as it corresponds to $\theta_{xm}$.

\begin{lstlisting}
R> bw_OLS <- COMMA_OLS(Mstar = bw_mstar[,1],
R>                     outcome = bw_y,
R>                     x_matrix = bw_x,
R>                     z_matrix = bw_z,
R>                     c_matrix = bw_c,
R>                     beta_start = bw_beta_start,
R>                     gamma_start = bw_gamma_start,
R>                     theta_start = bw_theta_start[-11],
R>                     tolerance = 1e-7,
R>                     max_em_iterations = 1500,
R>                     em_method = "squarem")
R> bw_OLS
R>    Parameter     Estimates Convergence Method
R> 1      beta1  -0.846812169        TRUE    OLS
R> 2      beta2   0.107762594        TRUE    OLS
R> 3      beta3   0.107837807        TRUE    OLS
R> 4      beta4 -18.245083179        TRUE    OLS
R> 5      beta5   0.773927649        TRUE    OLS
R> 6      beta6  -0.057283420        TRUE    OLS
R> 7      beta7   0.111048637        TRUE    OLS
R> 8      beta8   2.458154099        TRUE    OLS
R> 9      beta9   0.622722764        TRUE    OLS
R> 10   gamma11  -0.180421091        TRUE    OLS
R> 11   gamma21  -0.508468206        TRUE    OLS
R> 12   gamma31  -0.915139970        TRUE    OLS
R> 13   gamma41  -0.136061801        TRUE    OLS
R> 14   gamma12 -48.907580707        TRUE    OLS
R> 15   gamma22   0.765978153        TRUE    OLS
R> 16   gamma32 -16.887190012        TRUE    OLS
R> 17   gamma42   4.158003235        TRUE    OLS
R> 18    theta0   3.493664414        TRUE    OLS
R> 19   theta_m  -0.362627701        TRUE    OLS
R> 20   theta_x  -0.117757324        TRUE    OLS
R> 21  theta_c1   0.015685270        TRUE    OLS
R> 22  theta_c2  -0.503468498        TRUE    OLS
R> 23  theta_c3   0.088906401        TRUE    OLS
R> 24  theta_c4  -0.238715938        TRUE    OLS
R> 25  theta_c5  -0.015213574        TRUE    OLS
R> 26  theta_c6  -0.004364279        TRUE    OLS
R> 27  theta_c7   0.019881666        TRUE    OLS
\end{lstlisting}

Next, we use the \texttt{COMMA\_OLS\_bootstrap\_SE} function to obtain bootstrap standard errors for $(\boldsymbol{\beta}, \boldsymbol{\gamma}, \boldsymbol{\theta})$. Once again, arguments for the outcome distribution and interaction indicator are not required for this function.

\begin{lstlisting}
R> bw_OLS_SE <- COMMA_OLS_bootstrap_SE(parameter_estimates = 
R>                                       bw_OLS$Estimates,
R>                                     n_bootstrap = 100,
R>                                     n_parallel = 8,
R>                                     x_matrix = bw_x,
R>                                     z_matrix = bw_z,
R>                                     c_matrix = bw_c,
R>                                     tolerance = 1e-7,
R>                                     max_em_iterations = 1500,
R>                                     em_method = "squarem")
R> print(bw_OLS_SE$bootstrap_SE[,c("Parameter", "SE")], n = 27)
R> # A tibble: 27 x 2
R>    Parameter       SE
R>    <chr>        <dbl>
R>  1 beta1      0.240  
R>  2 beta2      0.247  
R>  3 beta3      0.0332 
R>  4 beta4      4.44   
R>  5 beta5      0.0698 
R>  6 beta6      0.212  
R>  7 beta7      0.0727 
R>  8 beta8      4.67   
R>  9 beta9      0.177  
R> 10 gamma11    0.205  
R> 11 gamma12   37.1    
R> 12 gamma21    0.178  
R> 13 gamma22    9.57   
R> 14 gamma31    0.331  
R> 15 gamma32   11.3    
R> 16 gamma41    0.0506 
R> 17 gamma42    3.32   
R> 18 theta0     0.0223 
R> 19 theta_c1   0.00722
R> 20 theta_c2   0.0534 
R> 21 theta_c3   0.00878
R> 22 theta_c4   0.0462 
R> 23 theta_c5   0.00449
R> 24 theta_c6   0.0784 
R> 25 theta_c7   0.0324 
R> 26 theta_m    0.0328 
R> 27 theta_x    0.0162
\end{lstlisting}

\end{appendices}

\end{document}